\documentclass[lettersize,journal]{IEEEtran}
\usepackage{amsmath,amsfonts}
\usepackage{algorithmic}
\usepackage{algorithm}
\usepackage{array}
\usepackage[caption=false,font=normalsize,labelfont=sf,textfont=sf]{subfig}
\usepackage{textcomp}
\usepackage{stfloats}
\usepackage{url}
\usepackage{verbatim}
\usepackage{graphicx}
\usepackage{cleveref}
\usepackage{cite}
\usepackage{mathtools}
\usepackage{pifont}
\usepackage{amsmath,amssymb,amsfonts}
\usepackage{amsthm,amsmath,amssymb}
\usepackage{mathrsfs}
\usepackage{enumerate}
\usepackage{times}
\usepackage{float}
\usepackage{color}
\usepackage{cuted}

\usepackage{xpatch}  
\usepackage{epstopdf}
\usepackage[utf8]{inputenc}
\usepackage[T1]{fontenc}

\makeatletter  
\xpatchcmd{\@thm}{\thm@headpunct{.}}{\thm@headpunct{}}{}{}
\makeatother  
\begin{document}

\title{Extended Target Adaptive Beamforming for ISAC: A Perspective of Predictive Error Ellipse}

\author{\IEEEauthorblockN{
    Shengcai Zhou, \emph{Student~Member, IEEE},
    Luping Xiang, \emph{Senior Member, IEEE},
    Yi Wang, \emph{Member, IEEE},\\
    Kun Yang, \emph{Fellow, IEEE},
    Kai Kit Wong, \emph{Fellow, IEEE},
    and Chan-Byoung Chae, \emph{Fellow, IEEE}}
    \vspace{-0.5 cm}\\

        \thanks{S. Zhou, L. Xiang and K. Yang are with the State Key Laboratory of Novel Software Technology, Nanjing University, Nanjing 210008, China, and School of Intelligent Software and Engineering, Nanjing University (Suzhou Campus), Suzhou 215163, China, email: zhshc@smail.nju.edu.cn, luping.xiang@nju.edu.cn, kunyang@nju.edu.cn.}
        \thanks{Y. Wang is with the Yangtze Delta Region Institute (Quzhou), University of Electronic Science and Technology of China, Quzhou 324003, China (e-mail: wangyi@csj.uestc.edu.cn).}
        \thanks{K.-K. Wong is with the Department of Electronic and Electrical Engineering, University College London, WC1E 7JE London, U.K., and also with the Yonsei Frontier Laboratory, Yonsei University, Seoul 03722, South Korea (e-mail: kai-kit.wong@ucl.ac.uk).}
        \thanks{C.-B. Chae is with the School of Integrated Technology, Yonsei University, Seoul 03722, South Korea (e-mail: cbchae@yonsei.ac.kr).}
	}

\maketitle



\maketitle

\begin{abstract}
Utilizing communication signals to extract motion parameters has emerged as a key direction in Vehicle-to-Everything (V2X) networks. Accurately modeling the relationship between communication signals and sensing performance is critical for the advancement of such systems. Unlike prior work that relies primarily on qualitative analysis, this paper derives the Cram\'{e}r-Rao Bound (CRB) for radar parameter estimation in the context of Orthogonal Frequency Division Multiplexing (OFDM) waveforms and Uniform Planar Array (UPA) configurations. Recognizing that vehicles may act as extended targets, we propose two New Radio (NR)-V2X-compatible beamforming schemes tailored to different phases of the communication process. During the initial beam establishment phase, we develop a beamforming approach based on the union of predictive error ellipses, which enhances scatterer localization through temporally assisted beam training. In the beam adjustment phase, we introduce an adaptive narrowest-beam strategy that leverages the positions of scatterers and the communication receiver (CR), enabling effective tracking with reduced complexity. The beam design problem is addressed using the minimum enclosing ellipse algorithm and tailored antenna control methods. Simulation results validate the proposed approach, showing up to a 32.4$\%$ improvement in achievable rate with a 32$\times$32 transmit antenna array and a 5.2$\%$ gain with an 8$\times$8 array, compared to conventional beam sweeping under identical SNR conditions.

\end{abstract}


\section{Introduction}

\subsection{Background}
\IEEEPARstart{N}{ext-generation} wireless communication technologies demand low-latency, high data rate, and high-precision positioning services in high-mobility scenarios \cite{10944644}. However, existing technologies, such as Dedicated Short Range Communication (DSRC), Cellular V2X (C-V2X), and Global Navigation Satellite Systems (GNSS), are unable to simultaneously meet these stringent requirements \cite{10502156}. On the other hand,  Integrated Sensing and Communication (ISAC) technology, by sharing spectrum, hardware platforms, baseband waveforms, and signal processing, allows for the simultaneous execution of sensing and communication functions \cite{9737357,10418473,10791445,peng2025simacsemanticdrivenintegratedmultimodal}. Therefore, ISAC is seen as one of the major emerging technologies for 6G \cite{10601686,10536135}.

With the assistance of ISAC technology, Vehicle-to-Everything (V2X) communication, which is typically deployed in high-mobility scenarios, has been the subject of many research efforts \cite{9171304,10061429,9947033,10502156,10806828,7990158,10845207}. In \cite{9171304}, Liu \textit{et al.} proposed a sensing-assisted downlink communication framework that utilizes an integrated waveform to acquire sensing information while transmitting communication data, thus reducing the overhead of pilots. Furthermore, Meng \textit{et al.}  and Du \textit{et al.}  conducted in-depth studies from the perspectives of complex roads \cite{10061429} and extended targets \cite{9947033}, proposing targeted beam tracking approaches. To quantify and optimize the reduction of pilots, Li \textit{et al.}  introduced a sensing-assisted New Radio (NR)-V2X frame structure \cite{10502156}. Additionally, based on the concept of hierarchical beam training \cite{7990158}, Zhou \textit{et al.} developed a temporal-assisted beamforming scheme \cite{10806828,10693589}. A recent work of Li \textit{et al.} have designed low-complexity beamforming algorithms \cite{10845207}. Despite the different perspectives, the above literature leverage massive Multiple-Input Multiple-Output (mMIMO) antennas and millimeter Wave (mmWave) technologies to achieve high beamforming gains, thereby promoting high-quality communication and sensing. Therefore, waveform design based on ISAC holds great promise. This paper investigates an adaptive beamforming scheme compatible with NR standards. 

\subsection{Related Work}
Some studies have focused on the separation of communication users and sensing targets \cite{10233699,9724205,9724174,9124713,10679658,10839626,9652071,8288677,10217169,9916163,10086626,10158711,10845891}. Li \textit{et al.} addressed the cross-layer trade-offs in time-division sensing and communication resource, assigning time slot to different targets during the sensing period and scheduling multiple users through a cycling protocol during the communication period \cite{10233699}. In addition, Liu \textit{et al.} examined joint transmission design for MIMO radar and downlink multi-user communication systems, focusing on the Signal to Interference plus Noise Ratio (SINR) balance in transmit beamforming and Dirty Paper Coding (DPC) mechanisms  \cite{9724205}. However, their approaches failed to unify the communication and radar waveforms. Hua \textit{et al.}  investigated integrated waveforms, considering an ISAC scenario involving a target and a multi-antenna user, with various sensing Cram\'{e}r-Rao Bound (CRB)-related metrics to illustrate the trade-off between CRB and data rate \cite{10217169}. Meanwhile, Wu \textit{et al.}  explored the design of a Quantized Constant-Envelope (QCE) integrated waveform for mMIMO ISAC systems, the waveform design problem was formulated to meet communication constraints based on constructive interference and QCE constraints \cite{10845891}. However, all the above-mentioned work do not account for the collaborative gains in ISAC.

Based on collaborative gains, research on sensing-assisted communication has been rapidly developed, with a focus on scenarios where the communication and sensing targets are the same \cite{9171304,10061429,9947033,10659350,zhou2025beamformingbasedachievableratemaximization,10502156,10330059}. Regarding target shape, Liu \textit{et al.} modeled vehicles as point targets and considers the downlink communication for the road side unit (RSU) serving vehicles \cite{9171304}. It proposed a sensing-assisted beam tracking method, which estimates the vehicle's motion parameters from ISAC echoes and then aligns the beam using Extended Kalman Filtering (EKF). Similarly, Meng \textit{et al.} modeled complex road geometries and proposed a curvilinear coordinate system with an Interacting Multiple Model (IMM)-based filtering scheme for vehicle tracking and identification \cite{10061429}. However, these works do not address extended targets. A sensing-assisted beam tracking method for extended vehicle was introduced by Du \textit{et al.}, who presented an ISAC-based dynamic beamforming scheme \cite{9947033}. Given that the Communication Receiver (CR) of the extended target cannot be tracked directly, the transmission beam needs to illuminate all scatterers, as partial illumination may not guarantee beam alignment with the CR. The dynamic beam scheme adjusts the beam according to the relative position of the vehicle and the RSU to ensure full vehicle coverage. To improve achievable rate, Du \textit{et al.} also proposed an ISAC-based alternating wide-narrow beam scheme. This is necessary, because when the vehicle is too close, the RSU beam become too wide, resulting in reduced achievable rate. The scheme reallocates resources for each slot, ensuring that the RSU first designs a wide beam to acquire accurate CR information and then transmitted a narrow beam to align with the CR. A similar idea was applied in UAV-to-vehicle communications \cite{10659350}. However, the accuracy of the tracking algorithm is limited by noise-imposed errors, and none of them account for situations where the beam is misaligned. In \cite{zhou2025beamformingbasedachievableratemaximization}, a uniform planar array (UPA) was used to consider error ellipses involving noise, providing explicit alignment probabilities to guide beam design. Furthermore, Li \textit{et al.} introduced a sensing-assisted NR-V2X frame structure that contrasts the differences from the conventional NR frame structure in three processes: initial access, connected mode, and beam beam failure and recovery \cite{10502156}. It quantified the advantages of using sensing to replace pilots. However, the aforementioned studies do not simultaneously consider the scenario involving the UPA, the extended target, and the NR-V2X frame structure, which is more representative of real-world conditions.

In the area of echo signal estimation,  Gonz\'{a}lez-Prelcic \textit{et al.} deployed additional radar equipment to assist with beam alignment, but does not evaluate the relationship between radar power and estimation accuracy \cite{7888145}. Liu \textit{et al.} reduced the need for extra radar hardware, yet it only develops a rough model based on the relationship between radar echo Signal-to-Noise Ratio (SNR) and motion parameter estimation errors \cite{9171304}. Similar studies include \cite{9246715,9947033,10659350}. In fact, evaluating the performance of estimators is challenging, as it is difficult to obtain closed-form solution for the classic metric, Mean Squared Error (MSE) \cite{8918497}. However, the CRB provides a lower bound for MSE. Many studies \cite{10806828,10061429,9945983,10543024}  employed generalized forms of the CRB for estimation, but did not derive a strictly application. The exact expression of the CRB depends on system configuration and signal structure, and it takes various forms \cite{9705498}. Xia \textit{et al.} derived the CRB for two-dimensional angles of arrival measurement, but this pertained to a bistatic manner and without considering the OFDM waveform \cite{10819949}. The CRBs mentioned in \cite{9705498} were all derived on the basis of radar waveforms. For example, \cite{9064480} derived the range-angle CRB for phased-array radar, while \cite{5672411} derived the angle and velocity CRB for the noncoherent moving colocated MIMO radar.  However, in colocated MIMO-OFDM systems, the coupling among multiple parameters makes the Fisher Information Matrix (FIM) analytically intractable for simplification or inversion. 

In the area of beam training, studies such as \cite{9171304,10061429,9947033,10659350} focused solely on the beam tracking phase, assuming perfect alignment during the beam training process, without considering beam training schemes based on ISAC. The NR-V2X frame structure proposed in \cite{10502156} demonstrates that ISAC can replace traditional pilots for beam alignment during the initial access. Its advantage lies in reducing the extensive beam scanning, enabling rapid alignment with the target. Additionally, Zhou \textit{et al.} proposed a beam alignment scheme assisted by temporal-assisted communication, drawing on the concept of hierarchical beam training \cite{10806828}. In this approach, the transmitter first emits a wide beam, which is then gradually narrowed into a directional beam as more precise information is gathered, achieving beam alignment. However, this scheme only considers point targets and generalized form of CRB.
\begin{table*}[t]
\centering
\caption{Contrasting Our Contributions To The State-Of-The-Art}
\setlength{\tabcolsep}{4mm}{
\begin{tabular}{l|c|c|c|c|c|c|c}
\hline 
Contributions & \textbf{this work} & \cite{9171304} & \cite{10061429} &\cite{9947033,10659350} & \cite{10502156} &\cite{zhou2025beamformingbasedachievableratemaximization} & \cite{10819949}
\\ \hline\hline
Integrated waveform & \ding{52} & \ding{51}  &  \ding{51} & \ding{51} & \ding{51} & \ding{51} &  \\
\hline
Extended target & \ding{52} &  &   & \ding{51} & & &  \\
\hline
Exact expression of CRB & \ding{52} &    &  &  && &\ding{51} \\
\hline
Error ellipse for beam alignment & \ding{52} &    & \ding{51}  &  & &\ding{51} & \\
\hline
Initial access based on ISAC & \ding{52} &    &  & &\ding{51} & \ding{51} & \\
\hline
\end{tabular}
}
\vspace{-0.3 cm}
\label{contributions}
\end{table*}

\subsection{Our Contributions}
Inspired by the aforementioned works, we propose a beam training and adaptive beam tracking scheme for extended targets based on the NR-V2X network. The innovations of this work are summarized in Table \ref{contributions}, with the main contributions outlined as follows:
\begin{itemize}
\item Based on the NR-V2X frame structure, we derive the CRB for vehicle motion parameters as a metric for sensing. We present the relationship between the echo SNR and sensing error as a strictly applicable closed-form solution. This approach contrasts with generalized qualitative expressions commonly used in \cite{10502156,9945983}.

\item We address the beam design for extended targets. Different from \cite{9947033}, we separate the beamforming process into the ISAC-based initial beam establishment (ISAC-IBE) and the ISAC-based adaptive beam adjustment (ISAC-ABA). This two-phase approach proves to reduce access delay and beam training overhead, while remaining compatible with the beam management procedures defined in the NR standard.

\item We also consider the impact of motion parameter measurement errors on beam design. The minimum external ellipse (MEE) algorithm is first adopted to address MEE problem, followed by the use of a mapping scheme to control the number of transmit antennas, thereby managing the beamforming process.

\item Compared to 5G NR beamforming scheme, the ISAC-IBE scheme achieves a balance between shorter expected access delay and higher sensing accuracy. In the beam adjustment phase, ISAC-ABA matches the ISAC-based regular beamforming (ISAC-RB) in achievable rate, yet requires only 1$\%$ of the algorithm calls with 8$\times$8  antennas. It also outperforms the beam sweeping scheme, improving the achievable rate by 5.2$\%$ with 8$\times$8  antennas, and by up to 32.4$\%$ with 32$\times$32 antennas. 
\end{itemize}
\begin{figure}[t] 
\centerline{\includegraphics[width=0.45\textwidth]{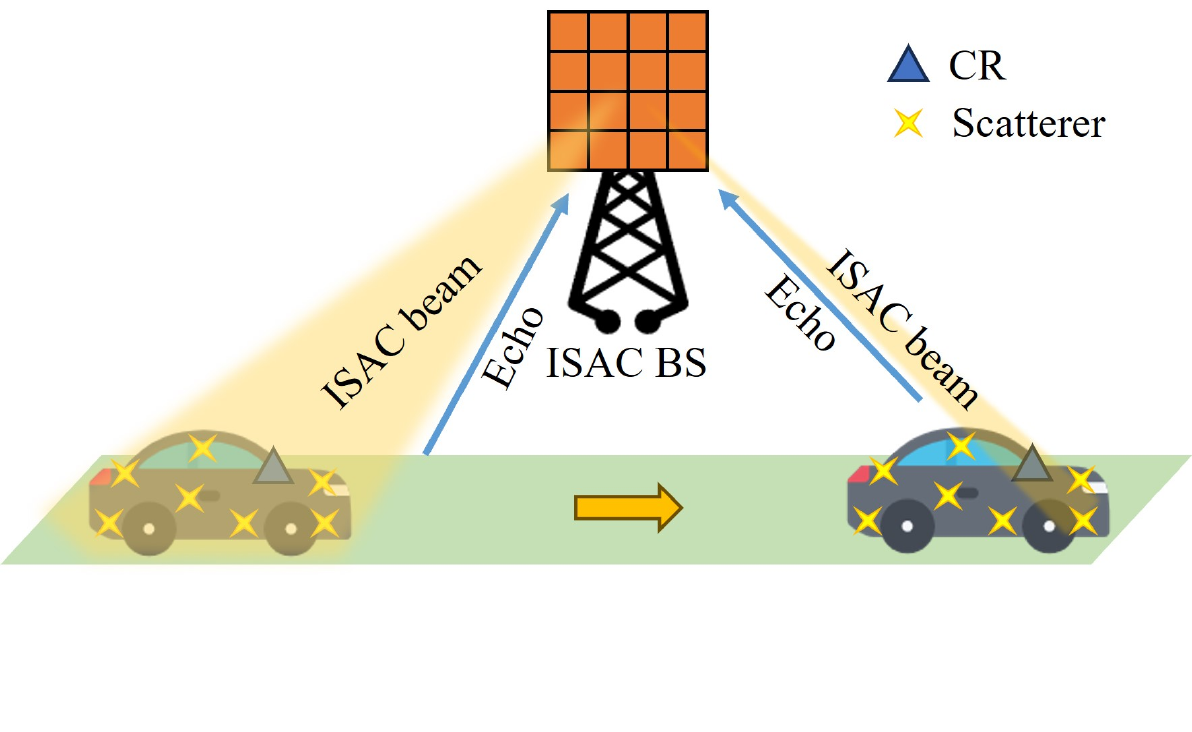}}
\vspace{-0.5 cm}
\caption{System model.}
\label{fig.scene}
\end{figure}
The remainder of this paper is organized as follows: Section II establishes the basic system model; Section III describes two communication modes based on the NR-V2X frame structure; Section IV presents the optimization problems and solves them; Section V shows the simulation analysis; and Section VI summarizes and prospects.

\textit{Notations:} In this paper, $(\cdot)^T$, $\left\| \cdot \right\|$, $(\cdot)^H$, $(\cdot)^{-1}$ and $\mathrm{diag}(\cdot)$ represent transpose, modulus of a vector, Hermitian, inverse  and diagonal matrix. \textbf{A} $\otimes$ \textbf{B} denotes the Kronecker product of two matrices \textbf{A} and \textbf{B}. \textbf{0} represents a vector where all elements are 0, while \textbf{I} represents identity matrix. $\mathfrak{R}$ denotes the real parts of a complex number. $\left[\cdot\right]_{ab}$ is defined as the element in the $a$ row and $b$ column of the matrix $\left[\cdot\right]$.

\section{System Model}
We consider a Base Station (BS) based on the ISAC NR-V2X protocol to provide communication services to the vehicle, as illustrated in Fig. \ref{fig.scene}. The BS is equipped with an mMIMO UPA that is both vertical and parallel to the roadside. The transmit and receive UPAs consist of $N_\text{z}$ rows and $N_\text{y}$ columns, i.e., the total number of antennas for each array is $ N_\text{t} = N_\text{r} = N_\text{z} \times N_\text{y}$. We assume that the isolation between the transmit and receive arrays is sufficiently adequate. Moreover, the BS operates in the mmWave frequency band, employing OFDM waveforms to transmit communication data and obtain sensing data from the reflected echoes. The corresponding channel is modeled as a narrowband mmWave channel. The vehicles travel along the straight road \footnote{The nonlinear road geometry will be taken into account in future work, and the initial results can be seen in \cite{10061429}.} and are modeled as extended targets, each equipped with a single antenna CR. According to \cite{9947033}, the vehicle is modeled as a volumetric scatterer whose internal reflections are spatially uniform in a statistical sense, and the CR will report its position during the uplink slot. When the BS has sufficiently accurate information about the scatterers, it can infer the relative position of the scatterers with respect to the CR and design the beam for the downlink slot accordingly. To maintain generality, we discretize the duration of interest into small slots, each with a duration of $\Delta T$. The beam is designed at each $\Delta T$ interval.
\subsection{Communication Model}
In the $n$th slot, the transmitted OFDM signal is represented as
\begin{align}
{s_n}\left( t \right) = \sum\limits_{k = 1}^{{N_{\text{sym}}}} {\sum\limits_{m = 1}^{{N_{\text{sub}}}} {{s_{n,m,k}}{e^{j2\pi \left( {m - 1} \right)\Delta ft}}{\rm{rect}}\left( {\frac{{t - \left( {k - 1} \right){T_\text{s}}}}{{{T_\text{s}}}}} \right)} },
\end{align}
where $\Delta f$ is the subcarrier interval, and $s_{n,m,k}$ denotes the data carried by the $k$th OFDM symbol of the $m$th subcarrier in slot $n$. ${T_\text{s}} = {T_{\text{cp}}} + T$ is the duration of the OFDM symbol, with $T_{\text{cp}}$ and $T$ being the cycle prefix duration and the elementary symbol duration, respectively. $N_{\text{sym}}$ and $N_{\text{sub}}$ represent the number of symbols and subcarriers per slot, respectively.

The steering vector of the transmit array is expressed as $\boldsymbol{a}(\theta ,\varphi)=\boldsymbol{a}_\text{z}(\theta ,\varphi) \otimes \boldsymbol{a}_\text{y}(\theta ,\varphi)$, where ${\theta}$ and ${\varphi}$ are the elevation and azimuth angles, and 
\begin{align}
\boldsymbol{a}_\text{z}(\theta ,\varphi)&={\left( {1,{e^{ - j\pi \sin  {\theta }}}, ... ,{e^{ - j\pi \left( {{N_\text{z}} - 1} \right)\sin  {\theta }}}} \right)^T},\\
\boldsymbol{a}_\text{y}(\theta ,\varphi)&={\left( {1,{e^{  j\pi \sin  {\varphi} \cos  {\theta} }}, ... ,{e^{  j\pi \left( {{N_\text{y}} - 1} \right)\sin  {\varphi} \cos  {\theta} }}} \right)^T},
\end{align}
are the transmit steering vectors in the horizontal and vertical directions, respectively. 

The impulse response of the narrowband mmWave channel is modeled as
\begin{align}
    \textbf{h}_n = {\alpha _{0,n}}{\boldsymbol{a}}({\theta _{0,n}},{\varphi _{0,n}}) + \sum\limits_{l = 1}^L {{\alpha _{l,n}}{\boldsymbol{a}}({\theta _{l,n}},{\varphi _{l,n}})},
\end{align}
where $\alpha$ denotes the path loss, and can be expressed as
\begin{align}
    \alpha_n = \frac{\lambda }{{4\pi  \cdot {d_n^\iota  }}},
\end{align}
with $\iota=0.95$  being the path loss exponent adopted from the empirical measurements in \cite{7109864} and $\lambda$ being the wavelength. $d_n$ denotes the distance between the CR and the BS. Besides, the channel include $L$ Non-Line-of-Sight (NLoS) paths, each of which can be considered as a MultiPath Component (MPC), with subjected to  Rayleigh distribution. Note that the power of the combined NLoS components $\sum\limits_{l = 1}^L {{\alpha _{l,n}}{\boldsymbol{a}}({\theta _{l,n}},{\varphi _{l,n}})}$ are much weaker (over 20dB \cite{7010533}) than that of the Line-of-Sight (LoS)  component ${\alpha _{0,n}}{\boldsymbol{a}}({\theta _{0,n}},{\varphi _{0,n}})$, so it is a reasonable assumption to consider only the LoS  component \footnote{When the LoS path is severely blocked, switching to the sub-6G band with conventional beam training can restore beam alignment \cite{10502156}. The modeling and analysis under NLoS conditions will be considered in future work.}, which becomes even more accurate as the beamwidth narrows. The channel therefore takes the following simplified form
\begin{align}
    \textbf{h}_n = {\alpha _{n}}{\boldsymbol{a}}({\theta _{n}},{\varphi _{n}}).
\end{align}
The received signal of the CR can be accordingly expressed as
\begin{align}
{c_n}\left( t \right) = \textbf{h}_n^H{\textbf{w}_n}{s_n}\left( {t - {\tau _{n}^\text{c}}} \right){e^{j2\pi {\mu _{n}^\text{c}}t}} + {z_\text{C}}\left( t \right),
\end{align}
where $\textbf{w}_n \in \mathbb{C}^{N_\text{t}}$ denotes beamforming vector, and $\tau _{n}^\text{c}$ and $\mu _{n}^\text{c}$ are time delay and the Doppler shift at the CR in the $n$th slot. $z_\text{C}(t) \sim\mathcal{CN}(0,\sigma_\text{C}^2)$ symbolizes the complex additive white Gaussian noise (AWGN) at the CR.

For simplicity, we consider only the data transmission portion of the communication frame, excluding other signals. Therefore, the achievable rate of the $n$th slot can be given by
\begin{align}
    {R_n} = {\log _2}\left( {1 + \text{SNR}_n^\text{C}} \right),
\end{align}
with $\text{SNR}_n^\text{C} = {\left| {{\textbf{h}_n^H}{\textbf{w}_n}} \right|}^2/\sigma_\text{C}^2$ being received SNR of the $n$th slot.

\subsection{Sensing Model}
\subsubsection{Echo Signal}
Based on the 5G NR standard \cite{3gpp38211_2022}, we consider the subcarrier spacing of 120kHz, with 14 OFDM symbols per slot, resulting in  $\Delta T = $ 0.125ms. Theoretically, the electromagnetic wave propagation distance within $\Delta T$ exceeds 10km, thus the echo data can be received and processed within $\Delta T$. For simplicity, self-interference is neglected, and it is assumed that the Doppler shift is solely caused by the motion of the vehicle. 

In the $n$th slot, the echo signal received by the BS from the extended vehicle can be formulated as
\begin{align}
{\textbf{r}_n}\left( t \right) = \sum\limits_{i = 1}^{{L_\text{s}}} {\beta _{i,n}}\textbf{b}({\theta _{i,n}},{\varphi _{i,n}}){{\boldsymbol{a}}^H}({\theta _{i,n}},{\varphi _{i,n}}){\textbf{w}_n}\nonumber  \\ 
\cdot{s_n}\left( {t - {\tau _{i,n}^\text{r}}} \right) {e^{j2\pi {\mu _{i,n}^\text{r}}t}} + {\textbf{z}_\text{R}}\left( t \right),
\end{align}
where $L_\text{s}$ is the number of scatterers, $\beta_{i,n}$ denotes reflection coefficient of the $i$th scatterer in the $n$th slot, and can be expressed as \cite{10529184}
\begin{align}
\beta_{i,n}= \frac{\lambda(G_\text{T} G_\text{R}\varepsilon_{i,n})^{1/2}}{(4\pi)^{3/2}{d_{i,n}^2}},
\end{align}
where $G_\text{T}$ and $G_\text{R}$ are the transmit and receive antenna gains, respectively. $\varepsilon_{i,n}$ and $d_{i,n}$ are the Radar Cross-Section (RCS) and the distance of the $i$th scatterer in the $n$th slot, respectively. Assume that $\varepsilon_{i,n}$ is the average of its distribution model \cite{mandelli2023survey} to simplify the analysis. In addition, ${\bf{b}}({\theta _{i,n}},{\varphi _{i,n}})$ is the steering vector of the receive array similar to ${\boldsymbol{a}}({\theta _{i,n}},{\varphi _{i,n}})$, and $\textbf{z}_\text{R}(t)$ symbolizes the AWGN. $\tau _{i,n}^\text{r}$ and $\mu _{i,n}^\text{r}$ are time delay and the Doppler shift of the $i$th scatterer in the $n$th slot.
\subsubsection{Sampling Process}
 Note that the time duration of an OFDM symbol is ${T_\text{s}}=\Delta T/14$, and the coherent time is ${T_\text{c}}=1/\mu^\text{r}$, where $\mu_{i,n}^\text{r}=2v_{i,n}^\text{rad}/\lambda$ , with $v_{i,n}^\text{rad}$ being the radial velocity. Therefore, we have ${T_\text{s}}\ll {T_\text{c}}$, indicating that inter-carrier interference can be ignored.  Sampling each OFDM symbol and applying blockwise Fourier Transform can discretize the signal and represent it in the frequency domain. In the $n$th slot, the discrete signal received at the $n_\text{r}$th antenna and the $k$th symbol can be represented as
\begin{align}
{\textbf{r}_{n_\text{r},k}} = \sum\limits_{i = 1}^{{L_\text{s}}} {{\beta _{i,n}}{{\left[ {\textbf{b}({\theta _{i,n}},{\varphi _{i,n}})} \right]}_{n_\text{r}}}{{\boldsymbol{a}}^H}({\theta _{i,n}},{\varphi _{i,n}}){\textbf{w}_n}}\\ \nonumber
{\cdot \left( {{\textbf{s}_k} \odot \boldsymbol{\eta} \left( {\tau _{i,n}^\text{r}} \right)} \right){\left[ {{\boldsymbol{\omega} ^\ast}\left( {\mu _{i,n}^\text{r}} \right)} \right]}_{n_\text{r}}}  + {\textbf{z}_{n_\text{r},k}}
\end{align}
where ${\textbf{r}_{n_\text{r},k}}=[{\text{r}_{n_\text{r},k}}[1],...,{\text{r}_{n_\text{r},k}}[{N_{\text{sub}}}]]^T$, $n_\text{r}=1,...,N_\text{r}$, ${\textbf{s}_k}=[s_{1,k},...,s_{N_\text{sub},k}]^T$, and
\begin{align}
    \boldsymbol{\eta} \left( {\tau _{i,n}^\text{r}}\right) &= {\left[ {1,{e^{ - j2\pi \Delta f{\tau _{i,n}^\text{r}}}},...,{e^{ - j2\pi \Delta f({N_{\text{sub}}}-1){\tau _{i,n}^\text{r}}}}} \right]^T},\\
    \boldsymbol{\omega} \left( {\mu _{i,n}^\text{r}} \right) &= {\left[ {1,{e^{ - j2\pi  {\mu _{i,n}^\text{r}} {T_\text{s}}}},...,{e^{ - j2\pi  {\mu _{i,n}^\text{r}}(N_{\text{sym}}-1) {T_\text{s}}}}} \right]^T}.
\end{align}
${\textbf{z}_{n_\text{r},k}}$ denotes the AWGN sample. The communication information can be removed through  element-wise division \cite{5776640}, after which the processed signal will only retain the sensing information. By aggregating $N_\text{sym}$ symbols in each slot, the signal received by the $n_\text{r}$th antenna can be written in a compact form as
\begin{align}\label{receivematrix}
\textbf{R}_{n_\text{r}} = \sum\limits_{i = 1}^{{L_\text{s}}} {\beta _{i,n}}{{\boldsymbol{a}}^H}({\theta _{i,n}},{\varphi _{i,n}}){\textbf{w}_n}\left[\textbf{b}({\theta _{i,n}},{\varphi _{i,n}})\right]_{n_\text{r}}  \nonumber\\ 
\cdot \boldsymbol{\eta} \left( {\tau _{i,n}^\text{r}} \right){\boldsymbol{\omega} ^H}\left( {\mu _{i,n}^\text{r}} \right)+ {\textbf{Z}_{n_\text{r}}},
\end{align}
where ${\textbf{Z}_{n_\text{r}}} \sim\mathcal{CN}(\textbf{0}_{N_\text{sub}\times N_\text{sym}},\sigma_\text{R}^2\textbf{I}_{N_\text{sub}\times N_\text{sym}})$. 
\subsubsection{Parameter Measurement}
Assume that the distance of the scatterers is greater than the distance resolution $\Delta r = \frac{c}{2B_\text{bw}}$, with $B_\text{bw}$ denoting the bandwidth. We employ the Two-dimensional Fast Fourier transform (2D-FFT) algorithm with respect to $\textbf{R}_{n_\text{r}}$, which performs IFFT in the subcarrier domain and FFT in the time domain, respectively. This process identifies $L_\text{s}$ peaks, with the row and column indices (${s_{i,n}^{\rm{d}}},{t_{i,n}^{\rm{d}}}$) corresponding to the measured distance and radial velocity of each scatterer, which can be respectively expressed as
\begin{align}
d_{i,n} &= \frac{{{s_{i,n}^{\rm{d}}}c}}{{2N_\text{sub}\Delta f}}, ~\text{and}\\
v^{\text{rad}}_{i,n} &= \frac{{{t_{i,n}^{\rm{d}}}c}}{{2{f_c}N_\text{sym}{T_{\rm{s}}}}},
\end{align}
where $f_c$ and $c$ denote the carrier frequency and the speed
of light, respectively. Then, the elevation and azimuth angles can be measured by using the classical MUltiple SIgnal classification (MUSIC) algorithm \cite{61541}. 

However, the above algorithms cannot evaluate the quality of parameter measurements, and thus cannot guide the beam design. To obtain closed-form solutions for the transmit power and the measurement errors, CRBs can be used as a performance metric. In the following, we derive the CRBs based on the echo signal. According to \cite{9947033}, the measurement errors can be regarded as Gaussian-distributed, which we express as
\begin{align}
{{\hat \theta }_{i,n}} &= {\theta _{i,n}} + {z_{{\theta _{i,n}}}},\label{meatheta}\\
{{\hat \varphi }_{i,n}} &= {\varphi _{i,n}} + {z_{{\varphi _{i,n}}}},\\
{{\hat \tau}_{i,n}^\text{r}} &= {\frac{2d_{i,n}}{c}} + {z_{{\tau_{i,n}}}},\\
{{\hat \mu}_{i,n}^\text{r}} &= {-\frac{2v_{i,n}\sin\varphi_{i,n}\cos\theta_{i,n}}{\lambda}} + {z_{{\mu_{i,n}}}},\label{meamu}
\end{align}
where ${z_{{\theta _{i,n}}}}$, ${z_{{\varphi _{i,n}}}}$, ${z_{{\tau_{i,n}}}}$ and ${z_{{\mu_{i,n}}}}$ signify additive noises with zero mean and variances ${\sigma^2_{{\theta _{i,n}}}}$, ${\sigma^2_{{\varphi _{i,n}}}}$, ${\sigma^2_{{\tau_{i,n}}}}$ and ${\sigma^2_{{\mu_{i,n}}}}$, respectively. 
\subsubsection{CRBs Derivation}
By vectorizing (\ref{receivematrix}), we get
\begin{align}\label{vectorEcho}
\textbf{r} = \sum\limits_{i = 1}^{{L_\text{s}}} {\beta _{i,n}}{{\boldsymbol{a}}^H}({\theta _{i,n}},{\varphi _{i,n}}){\textbf{w}_n}\textbf{b}({\theta _{i,n}},{\varphi _{i,n}}) \otimes {\boldsymbol{\omega} ^\ast}\left( {\mu _{i,n}^\text{r}} \right) \nonumber\\ 
\otimes \boldsymbol{\eta} \left( {\tau _{i,n}^\text{r}} \right)  + {\textbf{z}},
\end{align}
where $\textbf{z} \sim\mathcal{CN}(\textbf{0}_{N_\text{r}N_\text{sub}N_\text{sym}},\sigma_\text{R}^2\textbf{I}_{N_\text{r}N_\text{sub}N_\text{sym}})$. For convenience and without loss of generality, we omit the subscripts and only consider one scatterer.  (\ref{vectorEcho}) is then simplified to
\begin{align}
    \textbf{r} = \beta \left| {{{\boldsymbol{a}}^H}(\theta ,\varphi )\textbf{w}} \right|\textbf{e} + {\textbf{z}},
\end{align}
where $\textbf{e}=\textbf{b}({\theta },{\varphi}) \otimes {\boldsymbol{\omega} ^\ast}\left( {\mu ^\text{r}} \right)\otimes \boldsymbol{\eta} \left( {\tau^\text{r}} \right)$. The FIM for the complex Gaussian vector parameters is given by
\begin{align}
\textbf{J} = 2{\mathfrak{R}} \left\{ {{{\left( {\frac{{\partial \beta \left| {{{\boldsymbol{a}}^H}\textbf{w}} \right|\textbf{e}}}{{\partial \boldsymbol{\xi} }}} \right)}^H}{\bf{\Sigma}_\text{R} ^{ - 1}}\left( {\frac{{\partial \beta \left| {{{\boldsymbol{a}}^H}\textbf{w}} \right|\textbf{e}}}{{\partial \boldsymbol{\xi} }}} \right)} \right\},
\end{align}
where $\boldsymbol{\Sigma}_\text{R}  = {\sigma_\text{R} ^2}{\textbf{I}_{{N_\text{r}}{N_{\text{sub}}}{N_{\text{sym}}}}}$, $\boldsymbol{\xi}  = {\left[ {\boldsymbol{\alpha} \left| \boldsymbol{\beta}  \right.} \right]^T} = {\left[ {\theta ,\varphi ,\tau ,\mu \left| {\tilde \beta ,\bar \beta } \right.} \right]^T}$, where $\tilde \beta$ and $\bar \beta$ respectively represent the real part and the imaginary part of $\beta$. Let $\text{CRB}_{\boldsymbol{\alpha}}  = {\textbf{Q}^{ - 1}}$, the inverse of the FIM can be represented as
\begin{align}
{\textbf{J}^{ - 1}} = \left[ {\begin{array}{*{20}{c}}
{{\textbf{Q}^{ - 1}}}& \times \\
 \times & \times 
\end{array}} \right],
\end{align}
where $\times$ denotes the submatrix of no interest. The FIM is rewritten as a partitioned matrix containing $\boldsymbol{\alpha}$ and $\boldsymbol{\beta}$ as follows
\begin{align}
\textbf{J} = \left[ {\begin{array}{*{20}{c}}
{{\textbf{J}_{\boldsymbol{\alpha} \boldsymbol{\alpha} }}}&{{\textbf{J}_{\boldsymbol{\alpha} \boldsymbol{\beta} }}}\\
{\textbf{J}_{\boldsymbol{\alpha} \boldsymbol{\beta} }^T}&{{\textbf{J}_{\boldsymbol{\beta} \boldsymbol{\beta} }}}
\end{array}} \right].
\end{align}
According to the Schur complement \cite{5672411}, we get $\textbf{Q} = \left[ {{\textbf{J}_{\boldsymbol{\alpha} \boldsymbol{\alpha} }} - {{\textbf{J}_{\boldsymbol{\alpha} \boldsymbol{\beta} }}}{{\textbf{J}_{\boldsymbol{\beta} \boldsymbol{\beta} }^{-1}}}{{\textbf{J}_{\boldsymbol{\alpha} \boldsymbol{\beta} }}}^T} \right]$, expanding each block can yield (\ref{Q}), 
\begin{figure*}
\begin{align}\label{Q}
\textbf{Q} = 2R_\text{snr} \left[ {\begin{array}{*{20}{c}}
{{{\left\| {{\textbf{e}_\theta }} \right\|}^2}{{\sin }^2}{\Theta _1}}&{\frac{{\mathfrak{R} \left( {{\textbf{e}^H}{\textbf{G}_1}\textbf{e}} \right)}}{{{{\left\| \textbf{e} \right\|}^2}}}}&{\frac{{\mathfrak{R} \left( {{\textbf{e}^H}{\textbf{G}_2}\textbf{e}} \right)}}{{{{\left\| \textbf{e} \right\|}^2}}}}&{\frac{{\mathfrak{R} \left( {{\textbf{e}^H}{\textbf{G}_3}\textbf{e}} \right)}}{{{{\left\| \textbf{e} \right\|}^2}}}}\\
{\frac{{\mathfrak{R} \left( {{\textbf{e}^H}{\textbf{G}_1}\textbf{e}} \right)}}{{{{\left\| \textbf{e} \right\|}^2}}}}&{{{\left\| {{\textbf{e}_\varphi }} \right\|}^2}{{\sin }^2}{\Theta _2}}&{\frac{{\mathfrak{R} \left( {{\textbf{e}^H}{\textbf{G}_4}\textbf{e}} \right)}}{{{{\left\| \textbf{e} \right\|}^2}}}}&{\frac{{\mathfrak{R} \left( {{\textbf{e}^H}{\textbf{G}_5}\textbf{e}} \right)}}{{{{\left\| \textbf{e} \right\|}^2}}}}\\
{\frac{{\mathfrak{R} \left( {{\textbf{e}^H}{\textbf{G}_2}\textbf{e}} \right)}}{{{{\left\| \textbf{e} \right\|}^2}}}}&{\frac{{\mathfrak{R} \left( {{\textbf{e}^H}{\textbf{G}_4}\textbf{e}} \right)}}{{{{\left\| \textbf{e} \right\|}^2}}}}&{{{\left\| {{\textbf{e}_\tau }} \right\|}^2}{{\sin }^2}{\Theta _3}}&{\frac{{\mathfrak{R} \left( {{\textbf{e}^H}{\textbf{G}_6}\textbf{e}} \right)}}{{{{\left\| \textbf{e} \right\|}^2}}}}\\
{\frac{{\mathfrak{R} \left( {{\textbf{e}^H}{\textbf{G}_3}\textbf{e}} \right)}}{{{{\left\| \textbf{e} \right\|}^2}}}}&{\frac{{\mathfrak{R} \left( {{\textbf{e}^H}{\textbf{G}_5}\textbf{e}} \right)}}{{{{\left\| \textbf{e} \right\|}^2}}}}&{\frac{{\mathfrak{R} \left( {{\textbf{e}^H}{\textbf{G}_6}\textbf{e}} \right)}}{{{{\left\| \textbf{e} \right\|}^2}}}}&{{{\left\| {{\textbf{e}_\mu }} \right\|}^2}{{\sin }^2}{\Theta _4}}
\end{array}} \right]
\end{align}
\end{figure*}
where $R_\text{snr} = {{\left| \beta  \right|}^2}{{\left| {{{\boldsymbol{a}}^H}\textbf{w}} \right|}^2}/\sigma_\text{R} ^2$, ${\textbf{G}_1} = \left( {\textbf{e}_\theta ^H{\textbf{e}_\varphi }} \right)\textbf{I} - {\textbf{e}_\theta }\textbf{e}_\varphi ^H$, ${\textbf{G}_2} = \left( {\textbf{e}_\theta ^H{\textbf{e}_{\tau^\text{r}} }} \right)\textbf{I} - {\textbf{e}_\theta }\textbf{e}_{\tau^\text{r}} ^H$, ${\textbf{G}_3} = \left( {\textbf{e}_\theta ^H{\textbf{e}_{\mu^\text{r}} }} \right)\textbf{I} - {\textbf{e}_\theta }\textbf{e}_{\mu^\text{r}} ^H$, ${\textbf{G}_4} = \left( {\textbf{e}_\varphi ^H{\textbf{e}_{\tau^\text{r}} }} \right)\textbf{I} - {\textbf{e}_\varphi }\textbf{e}_{\tau^\text{r}} ^H$, ${\textbf{G}_5} = \left( {\textbf{e}_\varphi ^H{\textbf{e}_{\mu^\text{r}} }} \right)\textbf{I} - {\textbf{e}_{\mu^\text{r}} }\textbf{e}_\varphi ^H$, ${\textbf{G}_6} = \left( {\textbf{e}_{\tau^\text{r}} ^H{\textbf{e}_{\mu^\text{r}} }} \right)\textbf{I} - {\textbf{e}_{\tau^\text{r}} }\textbf{e}_{\mu^\text{r}} ^H$, and
\begin{equation}
\begin{split}
{\sin ^2}{\Theta _1} &= 1 - ({{{\mid{\textbf{e}_\theta ^H\textbf{e}}\mid }^2}}/{{{\parallel{{\textbf{e}_\theta }}\parallel }^2}{{\parallel\textbf{e}\parallel}^2}}),  \\
{\sin ^2}{\Theta _2} &= 1 - ({{{\mid{\textbf{e}_\varphi ^H\textbf{e}}\mid }^2}}/{{{\parallel{{\textbf{e}_\varphi }}\parallel }^2}{{\parallel\textbf{e}\parallel}^2}}),  \\ 
{\sin ^2}{\Theta _3} &= 1 - ({{{\mid{\textbf{e}_{\tau^\text{r}} ^H\textbf{e}}\mid }^2}}/{{{\parallel{{\textbf{e}_{\tau^\text{r}} }}\parallel }^2}{{\parallel\textbf{e}\parallel}^2}}), \\ 
{\sin ^2}{\Theta _4} &= 1 - ({{{\mid{\textbf{e}_{\mu^\text{r}} ^H\textbf{e}}\mid }^2}}/{{{\parallel{{\textbf{e}_{\mu^\text{r}} }}\parallel }^2}{{\parallel\textbf{e}\parallel}^2}}).
\end{split}
\end{equation}
Specifically, $\textbf{e}_\theta$, $\textbf{e}_\varphi$, $\textbf{e}_{\tau^\text{r}}$ and $\textbf{e}_{\mu^\text{r}}$ denote $\frac{{\partial \textbf{e}}}{{\partial \theta }}$, $\frac{{\partial \textbf{e}}}{{\partial \varphi }}$, $\frac{{\partial \textbf{e}}}{{\partial \tau^\text{r} }}$ and $\frac{{\partial \textbf{e}}}{{\partial v }}$, respectively. Let $\textbf{Q}=2R_\text{snr}\textbf{Q}'$, further expanding the derivative of $\textbf{Q}'$, we obtain (\ref{Qexpand}), where $N_\text{ss}=N_{{\text{sym}}}N_{{\text{sub}}}$, $\chi  = {N_\text{y}}\left( {N_\text{y}^2 - 1} \right)/12$, ${\chi _\tau } = {N_{{\text{sub}}}}\left( {N_{{\text{sub}}}^2 - 1} \right)/12$, and ${\chi _\mu } = {N_{{\text{sym}}}}\left( {N_{{\text{sym}}}^2 - 1} \right)/12$.
\begin{figure*}
\begin{align}\label{Qexpand}
\textbf{Q}' = \pi^2\left[ {\begin{array}{*{20}{c}}
{\left( {{{\sin }^2}\varphi {{\sin }^2}\theta {N_\text{z}} + {{\cos }^2}\theta {N_\text{y}}} \right)\chi {N_{{\text{ss}}}}}&{\sin \varphi \sin \theta \cos \varphi \cos \theta \chi {N_\text{z}}{N_{{\text{ss}}}}}&0&0\\
{\sin \varphi \sin \theta \cos \varphi \cos \theta \chi {N_\text{z}}{N_{{\text{ss}}}}}&{{{\cos }^2}\varphi {{\cos }^2}\theta \chi {N_\text{z}}{N_{{\text{ss}}}}}&0&0\\
0&0&{4\Delta {f^2}{\chi _\tau }{N_{\text{r}}}{N_{\text{sym}}}}&0\\
0&0&0&{4{T_s}^2{\chi _u}{N_{\text{r}}}{N_{\text{sub}}}}
\end{array}} \right]
\end{align}
\end{figure*}
Finally, the CRB of the parameters can be expressed respectively as
\begin{align}
&\text{CRB}_\theta  = {\left[ \textbf{Q}^{ - 1} \right]_{11}} = \frac{6}{R_\text{snr}{\pi ^2}\cos ^2\theta N_\text{y}^2\left( {N_\text{y}^2 - 1} \right)N_\text{ss}},\label{crbtheta}\\
&\text{CRB}_\varphi  = {\left[ {{\textbf{Q}^{ - 1}}} \right]_{22}} = \frac{{6\left( {{{\sin }^2}\varphi {{\tan }^2}\theta  + 1} \right)}}{R_\text{snr}{\pi ^2}{{\cos }^2}\varphi {{\cos }^2}\theta N_\text{y}^2\left( {N_\text{y}^2 - 1} \right){N_\text{ss}}},\label{crbphi}\\
&\text{CRB}_{\tau^\text{r}}  = {\left[ {{\textbf{Q}^{ - 1}}} \right]_{33}} = \frac{3}{{2R_\text{snr}{\pi ^2}\Delta {f^2}N_\text{y}^2{N_\text{ss}}\left( {N_\text{sub}^2 - 1} \right)}},\label{crbtau}\\
&\text{CRB}_{\mu^\text{r}}  = {\left[ {{\textbf{Q}^{ - 1}}} \right]_{44}} = \frac{3}{{2R_\text{snr}{\pi ^2}{T_\text{s}}^2N_\text{y}^2{N_\text{ss}}\left( {N_\text{sym}^2 - 1} \right)}},\label{crbmu}\\
&{\text{CR}}{{\text{B}}_{\theta ,\varphi }} = {\left[ {{\textbf{Q}^{ - 1}}} \right]_{12}} =   \frac{{-6\tan \varphi \tan \theta }}{{{R_\text{snr}}{\pi ^2}{{\cos }^2}\theta N_{\text{y}}^2\left( {N_{\text{y}}^2 - 1} \right){N_{{\text{ss}}}}}}.\label{crbthetaphi}
\end{align}
From (\ref{crbtheta}) to (\ref{crbthetaphi}), we conclude that all CRBs are influenced by the SNR, the symbol count, the subcarrier number, and the receiver antenna count. Besides, the  $\text{CRB}_\theta$ depends on the true elevation angle of the scatterer, the $\text{CRB}_\varphi$ is affected by both the true elevation and the azimuth angles, the $\text{CRB}_\tau$ is related to the subcarrier spacing, and the $\text{CRB}_\mu$ is influenced by the symbol duration.

\section{Two beamforming schemes based on NR-V2X frame structure}
\subsection{Overview}%
To facilitate wireless communication, traditional NR sets a large number of reference signals. Li \textit{et al.} proposed an NR-V2X frame structure \cite{10502156}, where during the initial access phase, the wide-angle ISAC signals are employed to monitor the  target's motion, thereby avoiding the latency introduced by extensive beam sweeping. In the connection phase, motion parameters are extracted from the echo of each  slot, and real-time beam tracking is performed using EKF, reducing the reliance on reference signals. Based on \cite{10502156}, in the initial beam establishment phase, we first transmit wide ISAC beam.  Then, using the echo information from each slot, we apply temporal-assisted beam design based on error ellipses to narrow the beamwidth until the BS has a sufficient understanding of the vehicle's shape. During the beam adjustment phase, the BS already has information about the relative positions of the CR and scatterers, meaning it only needs to track one scatterer to ensure reliable communication. We thus propose a real-time angle-domain narrowest beamforming, with the beam covering the nearest CR and scatterer.

\begin{figure}[t] 
\centerline{\includegraphics[width=0.38\textwidth]{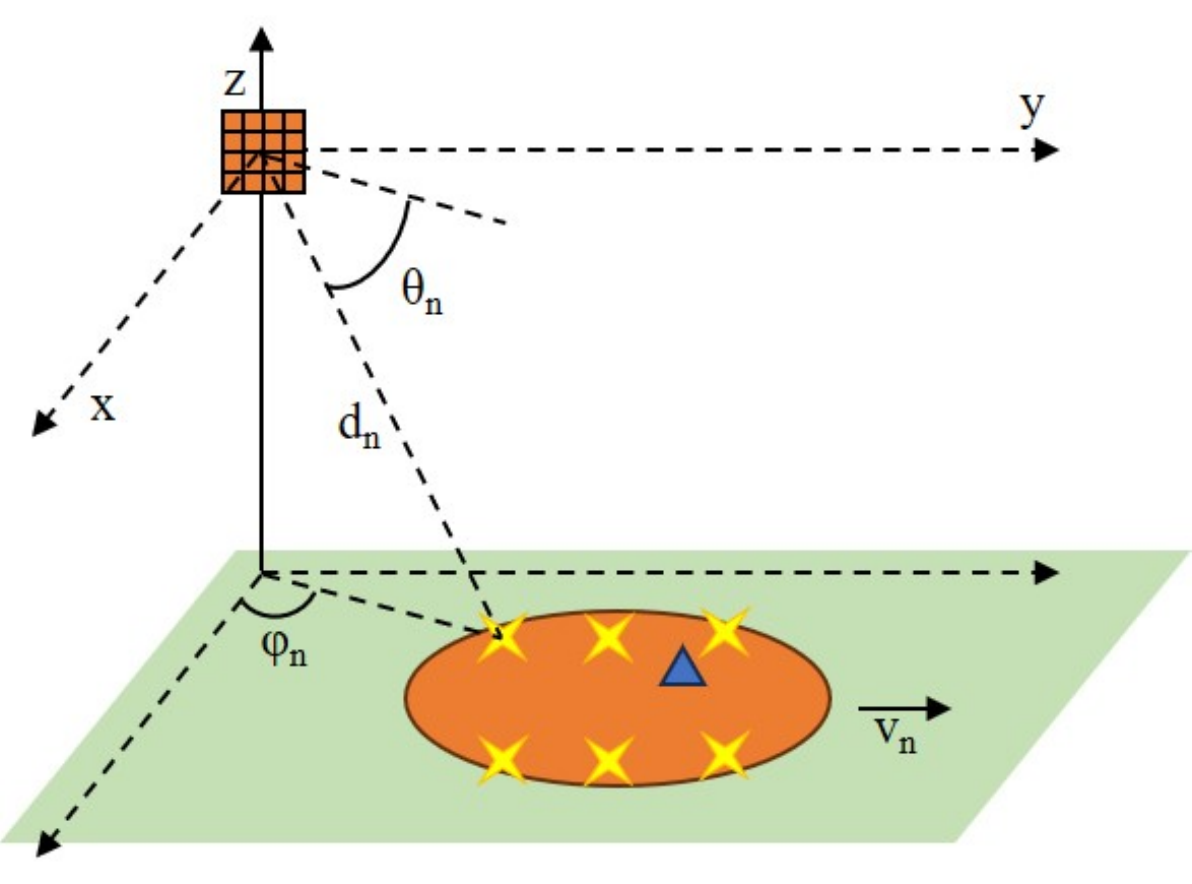}}
\caption{Cartesian coordinate system.}
\label{fig.coordinate}
\end{figure}

We model the vehicle with dimensions of 5 meters in length and 2 meters in width \footnote{The chosen vehicle dimensions are sufficiently representative \cite{NAP27236}, consistent with the assumption that the vehicle can be modeled as a uniformly distributed scatterer, thereby providing a reasonable approximation for simulating radar scattering.} \cite{9947033}. The kinematic parameters and geometric relationships relative to the BS are shown in Fig. \ref{fig.coordinate} (for one scatterer), where $v_n$ represents velocity and $h$ denotes the height of the UPA.
Using geometric relations and some approximations, the evolution model can be simply derived as
\begin{align}\label{evolution}
\left\{\begin{array}{l}
{\theta _{n + 1}} = {\theta _n} - d_n^{ - 1}{v_n}\Delta T\sin {\varphi _n}\sin {\theta _n}+ {\eta _\theta},\\
{\varphi _{n + 1}} = {\varphi _n} + d_n^{ - 1}{v_n}\Delta T\cos {\varphi _n}{\rm{sec}}{\theta _n}+ {\eta _\varphi},\\
{d_{n + 1}} = {d_n} + {v_n}\Delta T\cos {\theta _n}\sin {\varphi _n}+ {\eta _d},\\
{v_{n+1}} = {v_{n}} + {\eta _v},
\end{array} \right.
\end{align}
where $\eta _\theta$, $\eta _\varphi$, $\eta _d$ and $\eta _v$ denote the corresponding noises, which are generally  assumed to be zero-mean Gaussian distributed with variances of $\sigma^2 _\theta$, $\sigma^2 _\varphi$, $\sigma^2 _d$ and $\sigma^2 _v$, respectively. These noises are stemmed from approximation and other systematic errors.
\subsection{Initial Beam Establishment}
The BS determines whether a vehicle has entered the area of interest during the 1st slot using the wide ISAC beam, and subsequently establishes initial access in the 2nd slot. According to the sensing model of Section II, the measurements in $\theta-\varphi$ domain of each scatterer  follow the marginal Gaussian distribution, expressed as
\begin{align}
{\boldsymbol{\Psi} } = \left[ \begin{array}{l}
z_{\theta}\\
z_{\varphi}
\end{array} \right] \sim {\cal N}\left( {0,\bf{\Sigma} } \right),\bf{\Sigma}=\left[ \begin{array}{ll}
\text{CRB}_{\theta}&\text{CRB}_{\theta,\varphi}\\
\text{CRB}_{\theta,\varphi}&\text{CRB}_{\varphi}
\end{array} \right],
\end{align}
where ${\boldsymbol{\Psi}}$ denotes the uncertainty of angular measurement. Given a confidence level of 0.99, the error ellipse is obtained, with its major and minor axes represented by $2\sqrt {9.21{\lambda _\text{a}}} $ and $2\sqrt {9.21{\lambda _\text{b}}} $, respectively, from the cumulative Chi-square distribution. Note that $\lambda_\text{a}$ and $\lambda_\text{b}$ represent the two eigenvalues of $\bf{\Sigma}$. The angle of the elliptic direction is denoted by $\phi  = \arctan \left( {\frac{{{\boldsymbol{\nu} _1}\left( \varphi  \right)}}{{{\boldsymbol{\nu} _1}\left( \theta  \right)}}} \right)$, where $\boldsymbol{\nu} _1$ is the eigenvector of $\bf{\Sigma}$ corresponding to the larger eigenvalue.

\begin{figure}[t] 
\centerline{\includegraphics[width=0.47\textwidth]{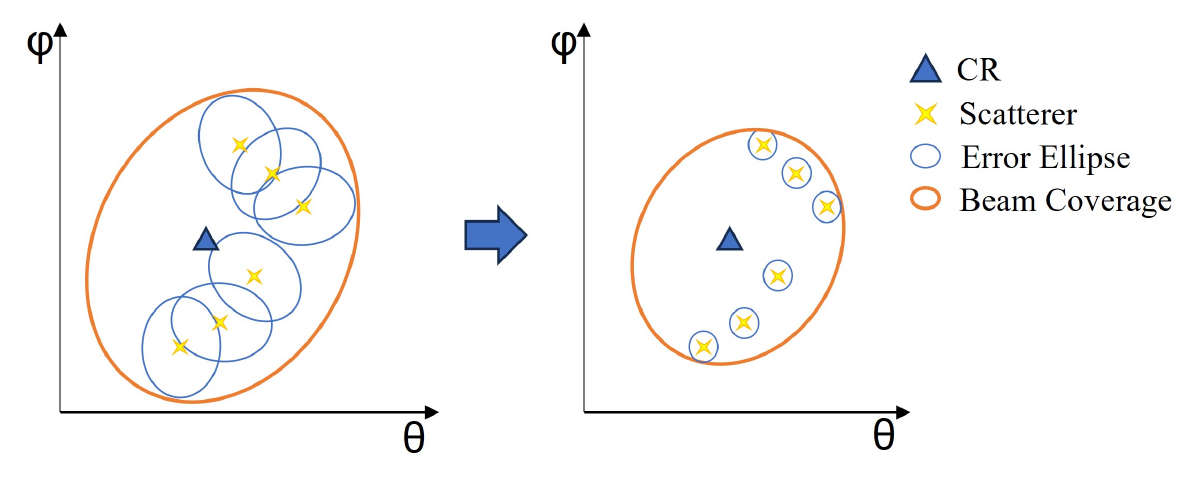}}
\vspace{-0.3 cm}
\caption{Beam aggregation process.}
\label{fig.aggregation}
\vspace{-0.3 cm}
\end{figure}
As illustrated in Fig. \ref{fig.aggregation}, the beam transmitted by the BS must cover the union of all error ellipses \footnote{Given that the scatterers are distinguishable, the union of error ellipses obtained from the single-scatterer angle-domain CRBs can serve as an approximation to the joint confidence ellipse of the multi-scatterer angle-domain CRB, for which a closed-form solution is generally intractable.}, in order to reduce the uncertainty associated with each scatterer. In each subsequent slot, the beam is narrowed based on the sensing information from its previous slot until convergence. This process results in sufficiently accurate position information for each scatterer. Furthermore, we exploit the uplink feedback specified in communication frame to report the CR position, thereby obtaining the CR’s relative position with respect to each scatterer. Based on the temporal-assisted beam design, we ensure that self-interference remains negligible (with each sensing occurring in just one slot), while significantly reducing the access delay and maintaining high connection quality. This beamforming scheme is called ISAC-IBE. 
\subsection{Beam Adjustment}
At this phase, the beam no longer needs to cover the entire vehicle, thereby enhancing the beam gain. However, two key issues must be addressed: \textit{1) The CR cannot be directly tracked. 2) Even if the CR can be tracked, in the face of potential beam interruptions, the BS must possess multi-step prediction capability for the CR's position.}

\begin{figure}[t] 
\centerline{\includegraphics[width=0.35\textwidth]{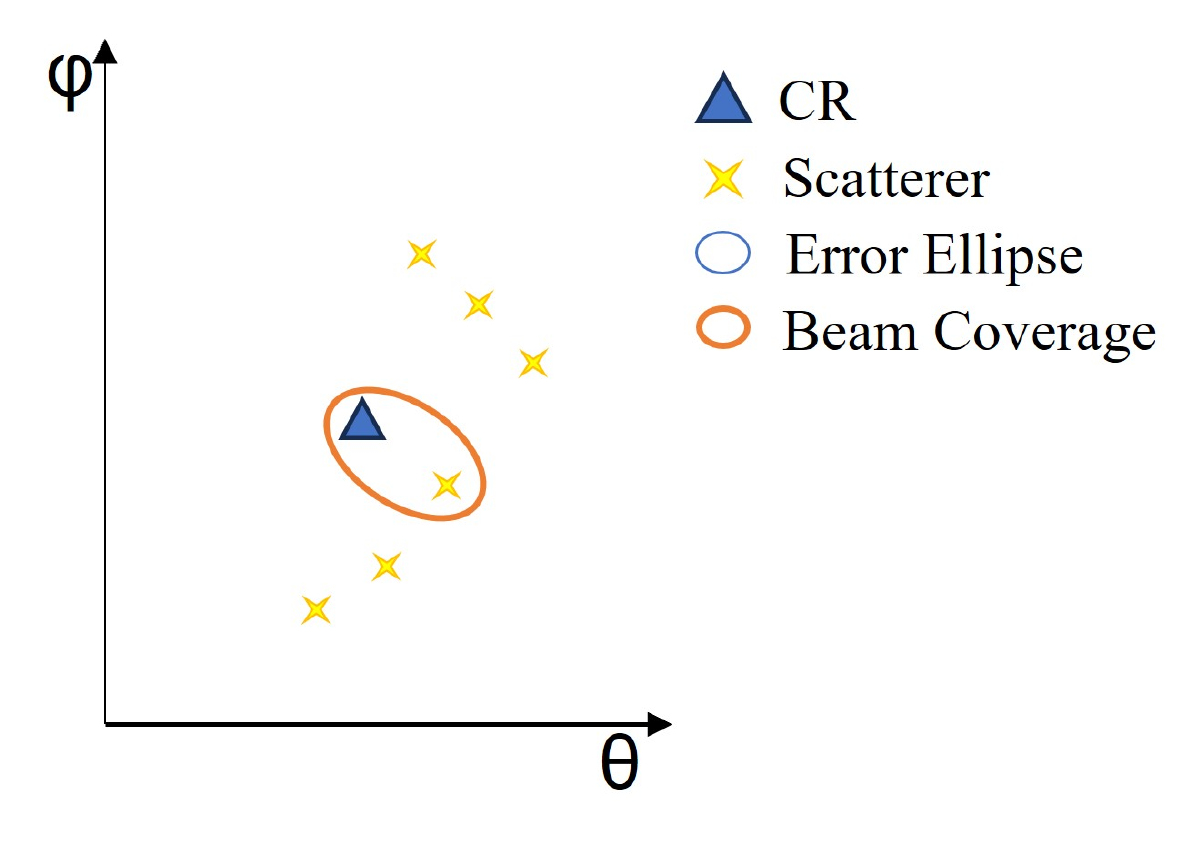}}
\vspace{-0.3 cm}
\caption{The narrowest beam.}
\label{fig.narrowestBeam}
\vspace{-0.3 cm}
\end{figure}
To cope with the first issue above, we consider a narrowest beam scheme, as illustrated in Fig. \ref{fig.narrowestBeam}, which simultaneously covers both the CR and the nearest scatterer. Through an indirect tracking approach, the measurement model for the CR can be derived as follows.

According to the scatterer measurement model (\ref{meatheta})-(\ref{meamu}), the Cartesian coordinates of the CR $(\hat x,\hat y)$ can be expressed as (ignoring subscripts)
\begin{align}
\hat x = \frac{{c\hat \tau^\text{r} \cos \hat \theta \cos \hat \varphi }}{2} + \Delta x,~~
\hat y = \frac{{c\hat \tau^\text{r} \cos \hat \theta \sin \hat \varphi }}{2} + \Delta y,
\end{align}
where $\Delta x$ and $\Delta y$ represent the relative coordinates from CR to scatterer. Note that $c\hat \tau^\text{r}/2=\hat d$. Next, The measurement models of the elevation angle $\theta ^{{\rm{CR}}}$, the azimuth angle $\varphi ^{{\rm{CR}}}$ and the distance $d ^{{\rm{CR}}}$ of CR, are expressed as
\begin{align}
\begin{split}
{\hat \theta ^{{\rm{CR}}}} &=  - \arctan \left( {\frac{h}{{\sqrt {{{\hat x}^2} + {{\hat y}^2}} }}} \right) = {\theta ^{{\rm{CR}}}} + {z_{{\theta ^{{\rm{CR}}}}}},\label{meacr}\\
{\hat \varphi ^{{\rm{CR}}}} &= \arctan \left( {\frac{{\hat y}}{{\hat x}}} \right) = {\varphi ^{{\rm{CR}}}} + {z_{{\varphi ^{{\rm{CR}}}}}},\\
{\hat d^{{\rm{CR}}}} &= \sqrt {{{\hat x}^2} + {{\hat y}^2} + {h^2}}  = {d^{{\rm{CR}}}} + {z_{{d^{{\rm{CR}}}}}},
\end{split}
\end{align}
where ${z_{{\theta ^{{\rm{CR}}}}}}$, ${z_{{\varphi ^{{\rm{CR}}}}}}$, and ${z_{{d^{{\rm{CR}}}}}}$ denote measurement noises. Due to the non-linearity of (\ref{meacr}), ${\hat \theta ^{{\rm{CR}}}}$, ${\hat \varphi ^{{\rm{CR}}}}$ and ${\hat d ^{{\rm{CR}}}}$ are not Gaussian. We adopt the first-order Taylor expansion to obtain the Gaussian approximation. Let ${\theta ^{{\rm{CR}}}} \buildrel \Delta \over = {g^\theta }\left( {\textbf{q}} \right)$, with ${\textbf{q}} = \left[ {\theta ,\varphi ,d} \right]$, we have
\begin{align}
{g^\theta }\left( \textbf{q} \right)   &\approx {g^\theta }\left( {\hat { \textbf{q}}} \right) + \frac{{\partial {g^\theta }}}{{\partial {{ \textbf{q}}^T}}}\left| {_{{ \textbf{q}} = \hat { \textbf{q}}}} \right. \cdot \left( {q - \hat { \textbf{q}}} \right)\nonumber\\
&= {g^\theta }\left( \hat{ \textbf{q}} \right) + {\textbf{G}^\theta }\left( {\textbf{q} - \hat{ \textbf{q}}} \right),\label{1ordertheta}
\end{align}
where ${\textbf{G}^\theta } = \left[ {\frac{{\partial {g^\theta }}}{{\partial \theta }},\frac{{\partial {g^\theta }}}{{\partial \varphi }},\frac{{\partial {g^\theta }}}{{\partial d}}} \right]_{{ \textbf{q}} = \hat { \textbf{q}}}$. Further, the elements of $\textbf{G}^\theta$ are derived as
\begin{align}
\begin{split}
\frac{{\partial {g^\theta }}}{{\partial \theta }} &=  - \frac{{hd\sin \theta \left( {X\cos \varphi  + Y\sin \varphi } \right)}}{{\left( {{X^2} + {Y^2} + {h^2}} \right)\sqrt {{X^2} + {Y^2}} }},\\
\frac{{\partial {g^\theta }}}{{\partial \varphi }} &= \frac{{hd\cos \theta \left( {Y\cos \varphi  - X\sin \varphi } \right)}}{{\left( {{X^2} + {Y^2} + {h^2}} \right)\sqrt {{X^2} + {Y^2}} }},\\
\frac{{\partial {g^\theta }}}{{\partial d}} &= \frac{{h\cos \theta \left( {X\cos \varphi  + Y\sin \varphi } \right)}}{{\left( {{X^2} + {Y^2} + {h^2}} \right)\sqrt {{X^2} + {Y^2}} }},
\end{split}
\end{align}
where $X = d\cos \theta \cos \varphi  + \Delta x$ and $Y = d\cos \theta \sin \varphi  + \Delta y$ for brief notations. Rewrite (\ref{1ordertheta}) as
\begin{align}
{g^\theta }\left( {\hat {\textbf{q}}} \right) \approx {g^\theta }\left( {\textbf{q}} \right) + {\textbf{G}^\theta }\left( {\hat {\textbf{q}} - {\textbf{q}}} \right),
\end{align}
with $\hat {\textbf{q}} \sim {\cal N}\left( {\textbf{q},{\bf{\Sigma} } ^{\rm{C}}} \right)$ and ${{\bf{\Sigma} } ^{\rm{C}}} = {\rm{diag}}\left( {{\bf{\Sigma} } ,{c^2}\text{CRB}_\tau /4} \right)$, we accordingly obtain ${g^\theta }\left( {\hat {\textbf{q}}} \right) \sim {\cal N}\left( {{g^\theta }\left( {\textbf{q}} \right),{\textbf{G}^\theta }{\bf{\Sigma} } ^{\rm{C}} {{\left( {{\textbf{G}^\theta }} \right)}^T}} \right)$, and ${z_{{\theta ^{{\rm{CR}}}}}}$ is approximated as zero-mean Gaussian distributed with variances of $\sigma _{{\theta ^{\text{CR}}}}^2 = {\textbf{G}^\theta }{\bf{\Sigma} } ^{\rm{C}} {\left( {{\textbf{G}^\theta }} \right)^T}$. By the same token, it can be derived that ${z_{{\varphi ^{{\rm{CR}}}}}}$ and ${z_{{d ^{{\rm{CR}}}}}}$ can be regarded as zero-mean Gaussian distributed with variance of $\sigma _{{\varphi ^{\text{CR}}}}^2 = {\textbf{G}^\varphi }{\bf{\Sigma} } ^{\rm{C}} {\left( {{\textbf{G}^\varphi }} \right)^T}$ and $\sigma _{{d ^{\text{CR}}}}^2 = {\textbf{G}^d }{\bf{\Sigma} } ^{\rm{C}} {\left( {{\textbf{G}^d }} \right)^T}$, respectively. The elements of $\textbf{G}^\varphi$ and $\textbf{G}^d$ are respectively summarized as
\begin{align}
\begin{split}
\frac{{\partial {g^\varphi }}}{{\partial \theta }} &= \frac{{d\sin \theta \left( {Y\cos \varphi  - X\sin \varphi } \right)}}{{{X^2} + {Y^2}}},\\
\frac{{\partial {g^\varphi }}}{{\partial \varphi }} &= \frac{{d\cos \theta \left( {Y\sin \varphi  + X\cos \varphi } \right)}}{{{X^2} + {Y^2}}},\\
\frac{{\partial {g^\varphi }}}{{\partial d}} &= \frac{{\cos \theta \left( {X\sin \varphi  - Y\cos \varphi } \right)}}{{{X^2} + {Y^2}}},
\end{split}
\end{align}
and
\begin{align}
\begin{split}
\frac{{\partial {g^d}}}{{\partial \theta }} &=  - \frac{{d\sin \theta \left( {X\cos \varphi  + Y\sin \varphi } \right)}}{{\sqrt {{X^2} + {Y^2} + {h^2}} }},\\
\frac{{\partial {g^d}}}{{\partial \varphi }} &= \frac{{d\cos \theta \left( {Y\cos \varphi  - X\sin \varphi } \right)}}{{\sqrt {{X^2} + {Y^2} + {h^2}} }},\\
\frac{{\partial {g^d}}}{{\partial {d_i}}} &= \frac{{\cos \theta \left( {Y\sin \varphi  + X\cos \varphi } \right)}}{{\sqrt {{X^2} + {Y^2} + {h^2}} }}.
\end{split}
\end{align}
Similarly, the velocity of CR are expressed as
\begin{align}
\hat v =  - \frac{{\lambda \hat {\mu^\text{r}} }}{{2\sin \hat \varphi \cos \hat \theta }} = v + {z_v}.
\end{align}
Let $v \buildrel \Delta \over = {g^v }\left( {\textbf{q}'} \right)$, with ${\textbf{q}'} = \left[ {\theta ,\varphi ,{\mu^\text{r}}} \right]$, we have
\begin{align}
{g^v}\left( \textbf{q}' \right) \approx {g^v}\left( {\hat {\textbf{q}}'} \right) + {G^v}\left( {{\textbf{q}'} - \hat {\textbf{q}}'} \right),
\end{align}
where ${\textbf{G}^v} = \left[ {\frac{{\partial {g^v}}}{{\partial \theta }},\frac{{\partial {g^v}}}{{\partial \varphi }},\frac{{\partial {g^v}}}{{\partial {\mu^\text{r}} }}} \right]_{{ \textbf{q}'} = \hat { \textbf{q}}'}$, and the elements of ${\textbf{G}^v}$ are
\begin{align}
\begin{split}
\frac{{\partial {g^v}}}{{\partial \theta }} &=  - \frac{{\lambda {\mu^\text{r}} \sin \theta }}{{2\sin \varphi {{\cos }^2}\theta }},\\ \frac{{\partial {g^v}}}{{\partial \varphi }} &= \frac{{\lambda {\mu^\text{r}} \cos \varphi }}{{2{{\sin }^2}\varphi \cos \theta }},\\ \frac{{\partial {g^v}}}{{\partial {\mu^\text{r}} }}& =  - \frac{\lambda }{{2\sin \varphi \cos \theta }}.
\end{split}
\end{align}
$z_v$ is seen as zero-mean Gaussian distributed with variance of $\sigma _{v}^2 = {\textbf{G}^v }{\bf{\Sigma} } ^{\rm{V}} {\left( {{\textbf{G}^v }} \right)^T}$, where ${\bf{\Sigma} } ^{\rm{V}}={\rm{diag}}\left( {{\bf{\Sigma} } ,\text{CRB}_{\mu^\text{r}}} \right)$.

In addressing issue 2, we can employ the classical EKF technique to predict the motion parameters of the CR. However, the EKF update requires the update of measurements, which in turn depends on the variation of the beam. The beam coverage can be formulated as a MEE problem, with further details provided in Section IV. Overall, continuous EKF tracking leads to significant computational overhead, whereas the evolution model inherently exhibits inherent robustness. Therefore, we propose an adaptive beamforming scheme. Assuming the beam is initially aligned with the CR, the BS updates the CR's parameters using the evolution model, directing the beam solely towards the CR. Due to the unmeasurable vehicle acceleration, the updated predictions of the CR is likely to deviate from the true values. By utilizing the uplink feedback, the BS can periodically obtain the true values of the vehicle parameters. Hence, when the angular estimation error exceeds $\Gamma_\text{r1}$, the EKF is triggered. The BS will then implement the narrowest beamforming scheme to cover an additional scatterer, adjusting the CR estimations until the error falls within the predefined threshold $\Gamma_\text{r2}$. Subsequently, the BS continues to direct the beam solely towards the CR. Repetition of the above process forms the adaptive adjustment of the beam. This beamforming scheme is called ISAC-ABA. 

\begin{figure}[t] 
\centerline{\includegraphics[width=0.4\textwidth]{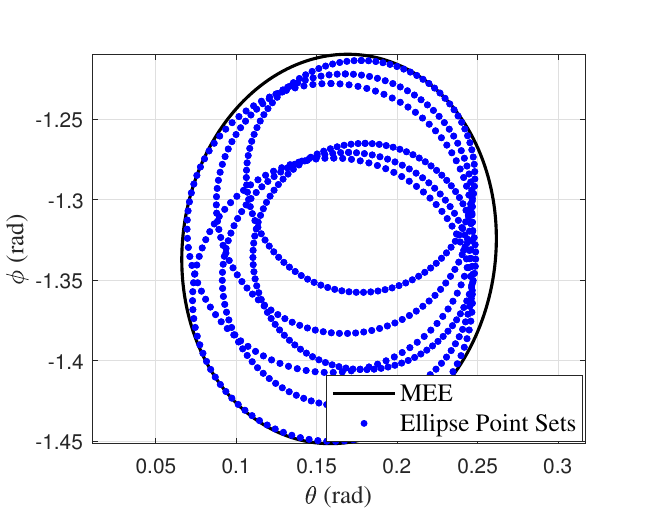}}
\caption{Simulation result of MEE algorithm.}
\label{fig.ellipseSetsShowCas}
\end{figure}

\section{Problem formulation and solution}
When the beam is required to encompass the union of the error ellipses of all scatterers, or simultaneously cover both the CR and the nearest scatterer, the underlying issue is to identify the MEE that satisfies the aforementioned conditions, which will define the beam coverage.
\subsection{Minimum External Ellipse Problem and Solution}
We first discretize each error ellipse as points in the Cartesian coordinate system, which can be expressed as ${\cal P} = \left\{ {{P_1},{P_2},...,{P_{{p_\text{num}}}}} \right\}$, with ${P_i} = \left( {{x_i^\text{E}},{y_i^\text{E}}} \right)$ and $p_\text{num}$ being the total number of discrete points. The MEE problem can be formulated as finding the smallest ellipse that encloses ${\cal P}$. The size of an MEE is defined by its area, expressed as $S_\text{area} = \pi ab$, where $a$ and $b$ represent the semi-major and semi-minor axes of the MEE, respectively. Since an MEE is arbitrary, its equation is modeled in central form
\begin{align}
A{\left( {x^\text{E} - {x^\text{E}_0}} \right)^2} + B\left( {x^\text{E} - {x_0^\text{E}}} \right)\left( {y - {y_0}} \right) + C{\left( {y^\text{E} - {y_0^\text{E}}} \right)^2} = F,\label{centralForm}
\end{align}
with $(x_0,y_0)$ being the center, and $A$,$B$,$C$ and $f$  being other parameters, respectively. Next, we transform (\ref{centralForm})  into a quadratic form as
\begin{align}
{\left[ {\begin{array}{*{20}{c}}
{x^\text{E} - {x_0^E}}\\
{y^\text{E} - {y_0^E}}
\end{array}} \right]^T}  \textbf{M}  \left[ {\begin{array}{*{20}{c}}
{x^\text{E} - {x^\text{E}_0}}\\
{y^\text{E} - {y^\text{E}_0}}
\end{array}} \right] = 1,
\end{align}
where $\textbf{M} = \left[ {\begin{array}{*{20}{c}}
{A/F}&{B/2F}\\
{B/2F}&{C/F}
\end{array}} \right]$.
For a real and symmetric matrix $\textbf{M}$, there exists an orthogonal matrix $\textbf{Q}$ such that $\textbf{Q}\textbf{M}{\textbf{Q}^T} = \left[ {\begin{array}{*{20}{c}}
{{\lambda _1}}&0\\
0&{{\lambda _2}}
\end{array}} \right]$, where ${\lambda _1}$ and ${\lambda _2}$ denote the eigenvalues of $\textbf{M}$. Therefore, we have ${a^2} = 1/{\lambda _1}$ and ${b^2} = 1/{\lambda _2}$, and get ${S_\text{area}} = \pi ab = \pi \sqrt {\frac{1}{{{\lambda _1}{\lambda _2}}}}  = \pi \sqrt {\frac{1}{{\det \left( \textbf{M} \right)}}} $. ${S_\text{area}}$ reaches its  minimum value when $\det \left( \textbf{M} \right)$ takes its maximum value. Consider $f$ as a normalized quantity greater than zero, the MEE problem can then be formulated as a convex optimization form
\begin{align}
&~~~~~~~~~~~~~~~\mathop {{\rm{min}}}\limits_{ A,B,C,{x^\text{E}_0},{y^\text{E}_0} }\frac{{4{F^2}}}{{4AC - {B^2}}}  \label{mee}\\
&~~~~~~~~~~~~~~~~~~~\;{\rm{s}}{\rm{.t}}{\rm{.}}~4AC - {B^2} > 0,\tag{\ref{mee}{a}}\\
&~~~~~~~~~~~~~~~~~~~~~~~~~A > 0,C > 0,\tag{\ref{mee}{b}}\\
&A{\left( {{x^\text{E}_i} - {x^\text{E}_0}} \right)^2} + B\left( {{x^\text{E}_i} - {x^\text{E}_0}} \right)\left( {{y^\text{E}_i} - {y^\text{E}_0}} \right) + C{\left( {{y^\text{E}_i} - {y^\text{E}_0}} \right)^2} \nonumber\\
&~~~~~~~~~~~~~~~~~~~~~~~~~~~~~~~~~~~~~~~~~~~~~~~~~~~~\le F,\forall i,\tag{\ref{mee}{c}}
\end{align}
where (\ref{mee}a) and (\ref{mee}b) are jointly derived from ellipse conditions ${\lambda _1}{\lambda _2} = \frac{{4AC - {B^2}}}{{4{F^2}}}$, ${\lambda _1} + {\lambda _2} = \frac{{A + C}}{F}$, and ${\lambda _1},{\lambda _2},F > 0$. (\ref{mee}c) ensures that $P_i$ is within the MEE. However, $F$ also affects the size of the MEE. We can add a convex optimization problem of the Minimum External Circle (MEC) in advance
\begin{align}
&~~~~~~~~~~~\mathop {{\rm{min}}}\limits_{ {x^\text{E}_1},{y^\text{E}_1} }r^2_1  \label{mec}\\
&~~~~~~~~~~~\;{\rm{s}}{\rm{.t}}{\rm{.}}~{\left( {{x^\text{E}_i} - {x^\text{E}_1}} \right)^2} + {\left( {{y^\text{E}_i} - {y^\text{E}_1}} \right)^2} \le r_1^2,\forall i,\tag{\ref{mec}{a}}
\end{align}
where $(x^\text{E}_1,y^\text{E}_1)$ and $r_1$ denote the center coordinate and radius of the MEC, respectively. (\ref{mec}a) ensures that $P_i$ is within MEC. Let the minimum radius be $r^*$, (\ref{mee}) can be optimized as
\begin{align}
&\mathop {{\rm{min}}}\limits_{ A,B,C,{x^\text{E}_0},{y^\text{E}_0} }\frac{{4{F^2}}}{{4AC - {B^2}}}  \label{mee1}\\
&~~~~~{\rm{s}}{\rm{.t}}{\rm{.}}~4AC - {B^2} > 4{F^2}{\left( {r^ * } \right)^4},\tag{\ref{mee1}{a}}\\
&~~~~~~~~~~(\text{\ref{mee}}{\mathrm{b}})~\text{and}~(\text{\ref{mee}}{\mathrm{c}}),\nonumber
\end{align}
where (\ref{mee1}a) represent the mathematical abstraction that the area of the MEE does not exceed that of the MEC, thereby ensuring a limited impact of $f$ on the MEE. All the above convex problems of (\ref{mee})-(\ref{mee1}) can be solved by calling the fmincon function in MATLAB\cite{fmincon2023}. The effect of the MEE algorithm is shown in Fig. \ref{fig.ellipseSetsShowCas}, where the MEE parameterized by the optimization results are shown to perfectly framing the ellipse point sets.

\begin{figure}[t] 
\centerline{\includegraphics[width=0.5\textwidth]{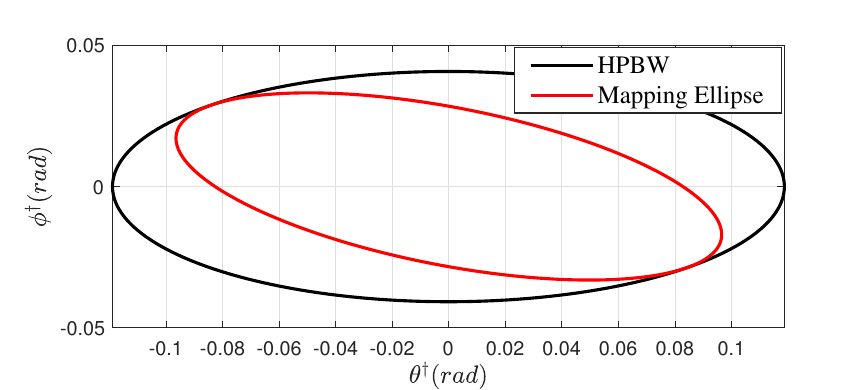}}
\caption{Simulation result of Mapping.}
\label{fig.mappingShowCase}
\end{figure}

\subsection{Antenna Control Scheme}
To make the beam cover the MEE, a simple and effective approach is to control the number of stimulated rectangular antennas \cite{10061429}, thereby regulating the Half-Power BeamWidth (HPBW). However, the relationship between the HPBW, determined by the number of antennas, and the required ellipse is not straightforward. To prevent the problem from becoming complicated, we propose an antenna control scheme based on mapping.

First, we map the HPBW to a standard ellipse in the ${\theta ^\dagger } - {\varphi ^\dagger }$ domain.  We omit the derivation and present the result as follows
\begin{align}
\left[ {\begin{array}{*{20}{c}}
{{\theta ^\dagger }}\\
{{\varphi ^\dagger }}
\end{array}} \right] = \left[ {\begin{array}{*{20}{c}}
{\cos \theta }&0\\
{\sin \theta \sin \varphi }&{ - \cos \theta \cos \varphi }
\end{array}} \right]\left[ {\begin{array}{*{20}{c}}
{\Delta \theta }\\
{\Delta \varphi }
\end{array}} \right],
\end{align}
where ($\theta,\varphi$) indicates the maximum gain direction, and ($\Delta \theta,\Delta \varphi$) indicates the the Euclidean distance from the boundary of the HPBW to ($\theta,\varphi$). In the ${\theta ^\dagger } - {\varphi ^\dagger }$ domain, by applying the same mapping to the MEE and then using the MEE algorithm to cover the MEE with the HPBW, the semi-major axis $\theta _{{\rm{BW}}}^\dagger$ and the semi-minor axis $\varphi _{{\rm{BW}}}^\dagger$ of the HPBW can be obtained, corresponding to the number of stimulated antennas ${N_\text{z}} = 1.78/\theta _{{\rm{BW}}}^\dagger$ and ${N_\text{y}} = 1.78/\varphi _{{\rm{BW}}}^\dagger$ respectively.  The effect of the mapping is shown in Fig. \ref{fig.mappingShowCase}, where the HPBW perfectly encompasses the mapping ellipse.

\section{Numerical Results}
This section presents the verification of the numerical results for the two beamforming schemes of ISAC-IBE and ISAC-ABA proposed in Section III, along with a complexity analysis. In the simulation, the vehicle moves from the left to the right, relative to the BS. The initial coordinates of the BS and  CR are provided as $\left[ {0,0,8} \right]$ and  $\left[ {10, - 38,1} \right]$, with six resolvable scatterers \footnote{Six scatterers are sufficient to capture the dominant scattering characteristics of the vehicle without introducing excessive modeling complexity.} uniformly distributed in the vehicle.  If not otherwise specified, main simulation parameters are listed in Table \ref{parameter}.

\begin{table}[h]
\centering
\renewcommand{\arraystretch}{1.2}
\caption{Simulation Parameters}
\begin{tabular}{c|c}
 \hline 
\textbf{parameter} & \textbf{value}  \\
\hline\hline
Transmit power $P_\text{T}$ (dBm) \cite{10061429} & 30   \\
\hline
Carrier frequency $f_c$ (GHz) \cite{10061429}& 30    \\
\hline
Number of subcarriers per slot $N_{\text{sub}}$  & 3300    \\
\hline
Noise power $\sigma_\text{C}^2$ or $\sigma_\text{R}^2$ (dBm) & -80   \\
\hline
Subcarrier spacing $\Delta f$ (kHz) & 120  \\
\hline
Slots per frame $N_{\text{slot}}$ (dB)& 80   \\
\hline
Bandwidth $B_\text{bw}$ (MHz) & 400   \\
\hline
Receive antennas $N_{\text{r}}$ (MHz) & 64   \\
\hline
Noise of evolution elevation angle $\sigma^2 _\theta$ (degree) & 0.02   \\
\hline
Noise of evolution azimuth angle $\sigma^2 _\varphi$ (degree) & 0.02   \\
\hline
Noise of evolution distance $\sigma^2 _d$ (m) & 0.2  \\
\hline
Noise of evolution velocity $\sigma^2 _v$ (m/s) & 0.25  \\
\hline
Vehicle length (m) & 5  \\
\hline
Vehicle width (m) & 2  \\
\hline
Number of OFDM symbols per slot  & 14  \\
\hline
Slot duration $\Delta T$ (ms)  & 0.125  \\
\hline
Number of scatterers  & 6  \\
\hline
Distance resolution (m)  & 0.375  \\
\hline
\end{tabular}
\label{parameter}
\end{table}

\subsection{Performance of Initial Beam Establishment}
\begin{figure}[h] 
    \centerline{\includegraphics[width=0.3\textwidth]{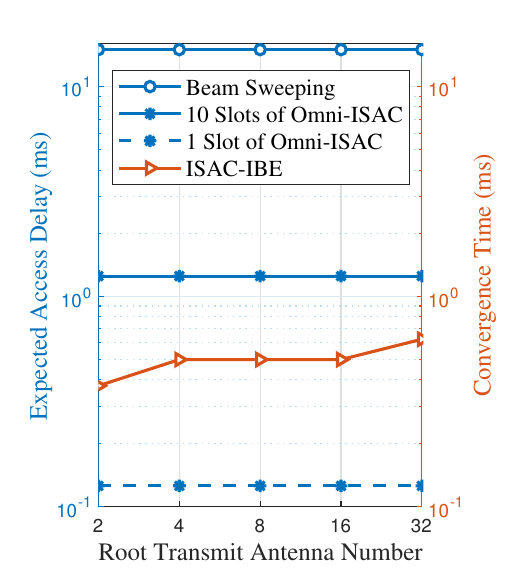}}
    \caption{Convergence time of ISAC-IBE and expected access delay of other schemes.}
    \label{fig.convergence}
\end{figure}
In this subsection, we evaluate the superiority of our ISAC-IBE by conducting a comparative analysis with the beam sweeping scheme and the Omni-ISAC scheme \cite{10502156}. The beam sweeping scheme involves transmitting 64 Synchronization Signal Block (SSB) beams every 20ms, with the receiver selecting the beam with the maximum reception power and providing feedback to the transmitter. The Omni-ISAC scheme refers to directly sensing the receiver's position through the echo of the wide-angle ISAC signal from the previous slots, from which we investigated the cases of sensing through the previous 1 slot and 10 slots. Fig. \ref{fig.convergence} first presents the access delay of the comparative schemes, with the values derived from \cite{10502156}. As indicated in Section III, the access delay of ISAC-IBE is equivalent to that of 1 slot of Omni-ISAC. It then shows the convergence time of ISAC-IBE, which increases with the number of transmit antennas, as ISAC-IBE leverages array gain to further enhance sensing accuracy.
\begin{figure*}[th]
    \centering
    \subfloat[]{\includegraphics[width=2.2in]{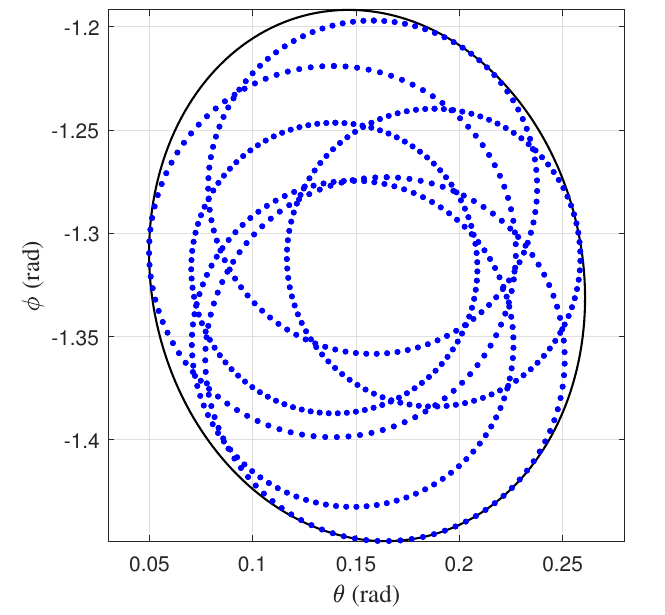}%
        \label{wnbeam1}}
    \hfil
    \subfloat[]{\includegraphics[width=2.2in]{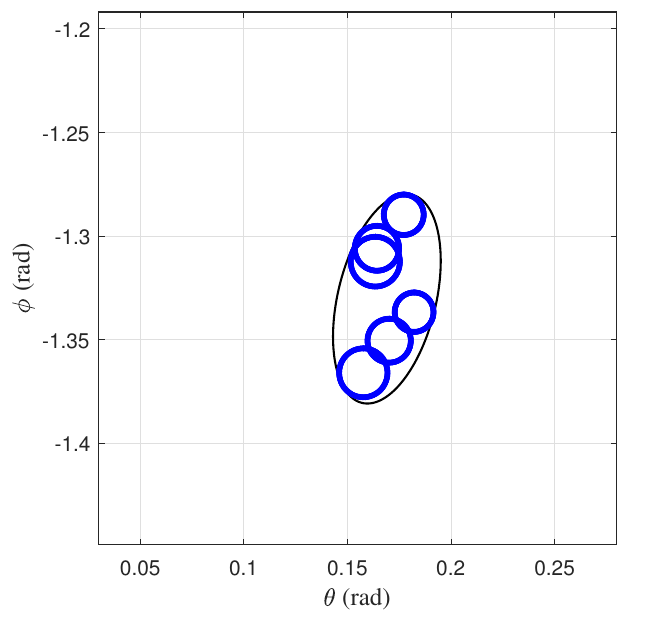}%
        \label{wnbeam2}}
    \hfil
    \subfloat[]{\includegraphics[width=2.2in]{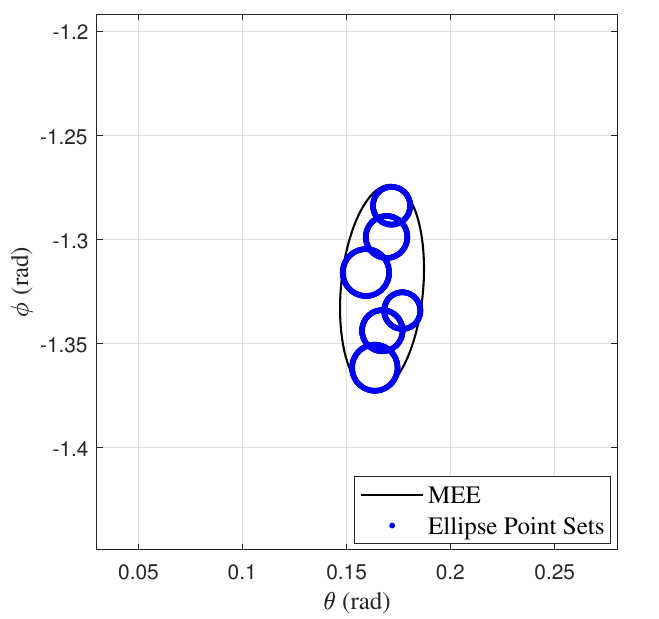}%
        \label{wnbeam3}}
    \caption{The variations of the error ellipse and MEE under different slots. (a) slot 1. (b) slot 2. (c) slot 3.}
    \label{wnbeam}
\end{figure*}

\begin{figure}[h] 
    \centerline{\includegraphics[width=0.52\textwidth]{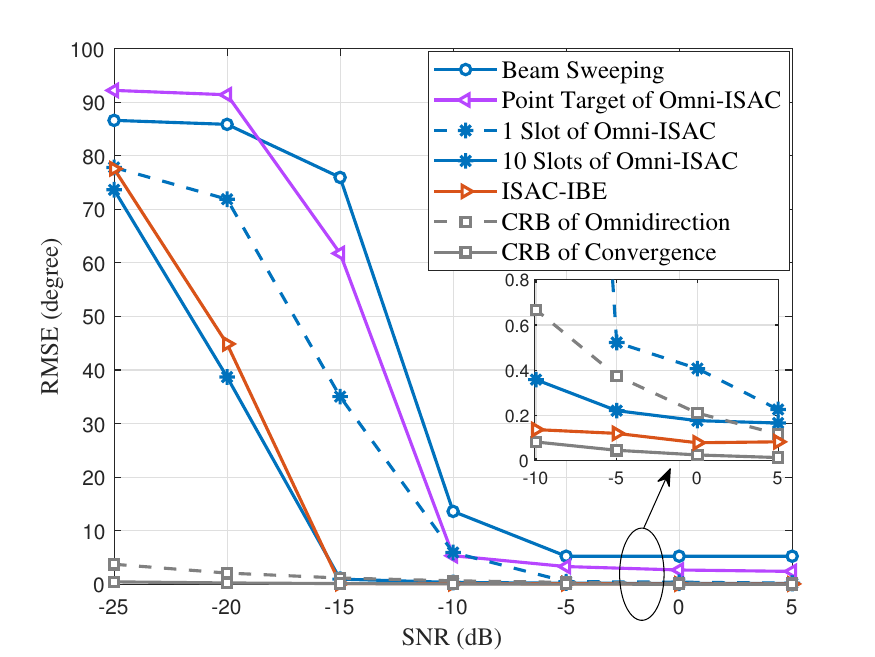}}
    \caption{RMSE of azimuth angle estimation for different schemes under different SNR.}
    \label{fig.RmseRMSE}
\end{figure}
In Fig. \ref{fig.RmseRMSE}, we introduce several comparative schemes to illustrate the root mean square error (RMSE) of azimuth angle estimation under different SNR. ``Point-target of Omni-ISAC" treats the vehicle as a point target and senses a randomly scatterer in 1 slot. ``Omni-CRB" and ``Convergence CRB" represent the best theoretical estimates obtained from ``1 slot of Omni-ISAC" and the converged sensing of ISAC-IBE, respectively. For beam sweeping and ``Point-target of Omni-ISAC", the RMSE remains significant even at high SNR. In beam sweeping, the codebook is fixed, which implies that the estimable angle is also fixed. In the case of ``Point-target of Omni-ISAC", treating a random scatterer as the CR prevents obtaining comprehensive knowledge of the vehicle. Additionally, we observe that the ``Omni-CRB" curve lies below the ``1 slot of Omni-ISAC" curve, and the ``CRB of Convergence" curve lies beneath the ISAC-IBE curve, validating the correctness of CRB as a lower bound.

Next, we present the convergence process of ISAC-IBE's MME in the $\theta-\varphi$ domain in Fig. \ref{wnbeam}, where the blue dot set represents the error ellipses for each scatterer. As the number of slots increases, each error ellipse shrinks, and the are of the MME decreases. This demonstrates the improvement in sensing accuracy brought by ISAC-IBE, corroborating Fig. \ref{fig.aggregation}.

\subsection{Performance of Beam Adjustment}
\begin{figure}[t] 
    \centerline{\includegraphics[width=0.48\textwidth]{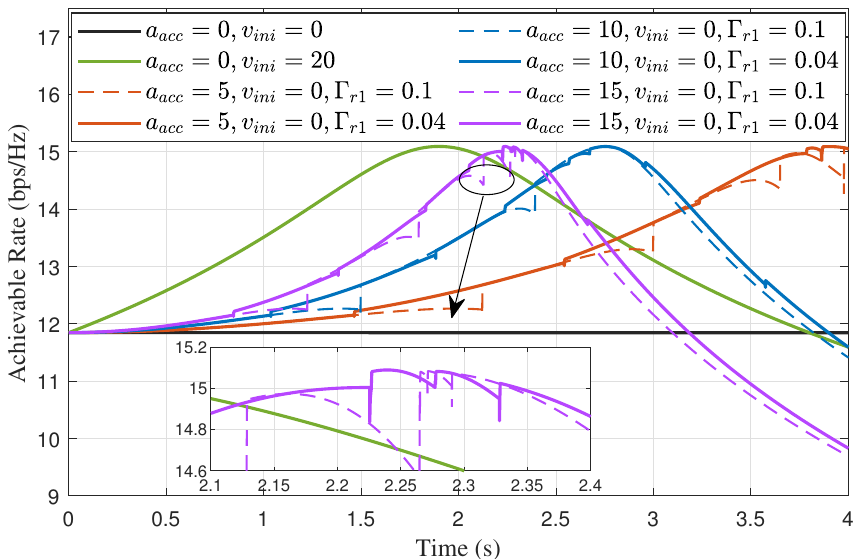}}
    \caption{Comparison of achievable rate under different acceleration, initial velocity, and $\Gamma_\text{r1}$.}
    \label{fig.accInitialV}
\end{figure}

The simulation setup considers a vehicle that moves from one side of the BS, passes through it, and reaches the opposite side. We first evaluate the variation in achievable rate of the ISAC-ABA scheme under different acceleration $a_\text{acc}$, initial velocity $v_\text{ini}$, and angle estimation error threshold $\Gamma_\text{r1}$ in Fig. \ref{fig.accInitialV}. Taking the stationary vehicle as a reference, the curve with $a_\text{acc}=0\text{m/s}^2,v_\text{ini}=20\text{m/s}$ illustrates the variation in achievable rate as the distance between the vehicle and the BS changes. Additionally, the role of the evolution model is evident. When $v_\text{ini}$ remains constant, the sensing function is not activated. With $v_\text{ini}=0$ and $a_\text{acc}$ of 5, 10, and 15, it can be observed that a higher acceleration leads to more frequent sensing activations, and the closer the vehicle is to the BS, the more frequent the sensing calls become. Comparing $\Gamma_\text{r1}$ of 0.1rad and 0.04rad, a larger tolerable angular error results in fewer sensing activations, thereby reducing the overall achievable rate.

\begin{figure}[h]
    \centering
    \subfloat[]{\includegraphics[width=3.7in]{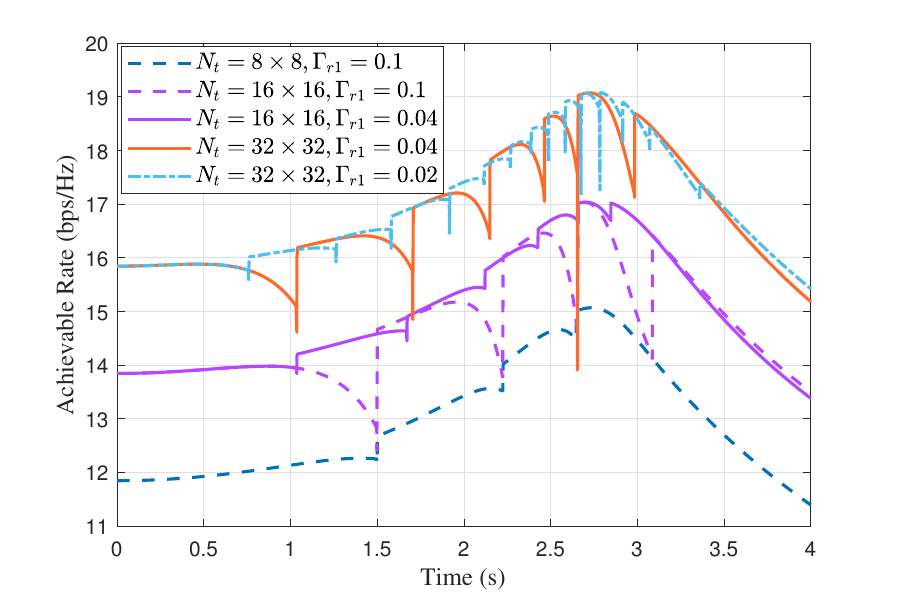}%
        \label{antenna4con}}
    \hfil
    \subfloat[]{\includegraphics[width=3.7in]{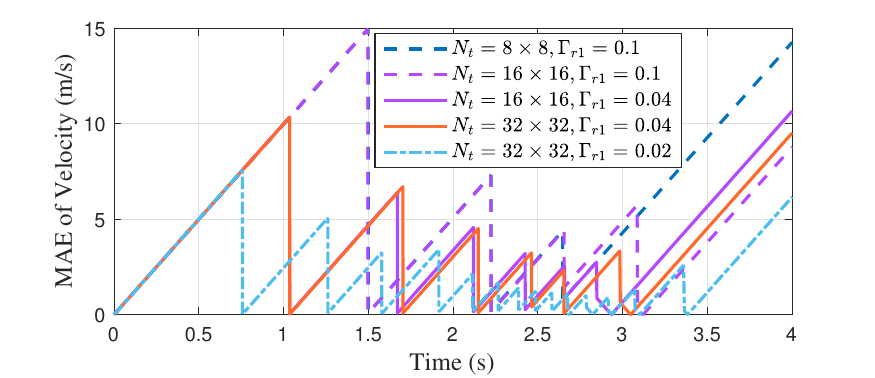}%
        \label{antenna4conV}}
    \caption{Comparison of the performance of ISAC-ABA under different transmit antennas and $\Gamma_\text{r1}$. (a) achievable rate. (b) MAE of velocity.}
    \label{antenna4c}
\end{figure}
Fig. \ref{antenna4c}(a) illustrates the variation in achievable rate of ISAC-ABA under different numbers of transmit antennas and $\Gamma_\text{r1}$. With $a_\text{acc}=10\text{m/s}^2,v_\text{ini}=0\text{m/s}$, as the number of transmit antennas increases, the array gain improves, leading to a higher achievable rate. Notably, under the same $\Gamma_\text{r1}$, a larger number of transmit antennas results in a greater reduction in achievable rate before sensing is activated, due to the decrease in HPBW. Therefore, it is necessary to lower $\Gamma_\text{r1}$ as the number of transmit antennas increases. Fig. \ref{antenna4c}(b) presents the corresponding Mean Absolute Error (MAE) in velocity. As sensing calls become more frequent, the MAE decreases. When $\Gamma_\text{r1}$ is not met, the velocity estimate remains at the value from the previous sensing activation.

\begin{figure}[h]
\centering
\subfloat[]{\includegraphics[width=3.7in]{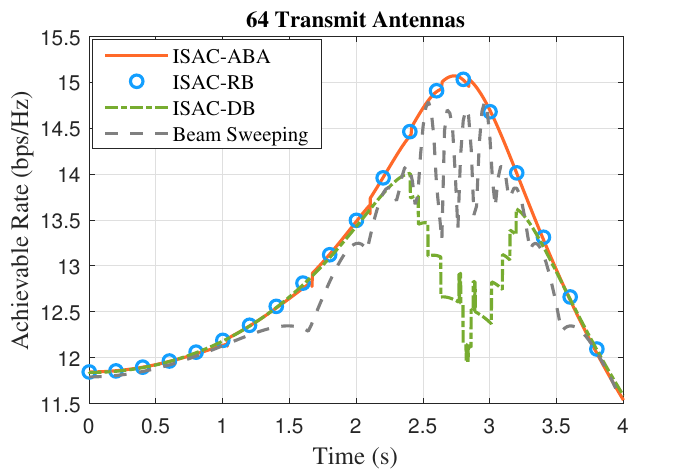}%
\label{adaptiveVSother8new}}
\hfil
\subfloat[]{\includegraphics[width=3.7in]{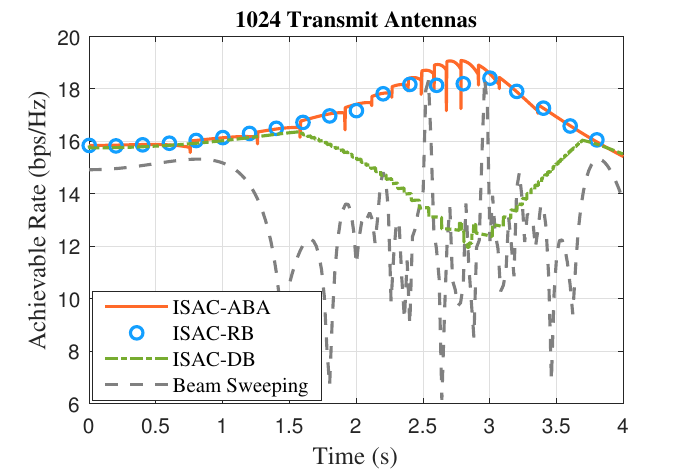}%
\label{adaptiveVSother32new}}
\caption{Comparison of the achievable rate of beam schemes under different transmit antennas. (a) 64 transmit antennas. (b) 1024 transmit antennas.}
\label{adaptiveVSothernew}
\end{figure}
Next, we examine the superiority of our ISAC-ABA, as shown in Fig. \ref{adaptiveVSothernew}, by adding three comparison groups:
\begin{itemize}
    \item The regular beamforming scheme based on ISAC (ISAC-RB), which disregards the triggering threshold, is equivalent to ISAC-ABA with continuous sensing throughout the process.
    \item The ISAC-DB scheme \cite{9947033} involves the BS continuously covering the entire vehicle with wide beam, eliminating the need for EKF activation.
    \item The beam sweeping scheme \cite{10502156} adjusts the beam by transmitting a subset of 8 SSB beams every 20ms, rather than 64 SSB beams.
\end{itemize}
In Fig. \ref{adaptiveVSothernew}(a), with 64 transmit antennas and $\Gamma_\text{r1}$ set to 0.04 rad, the achievable rates of ISAC-ABA and ISAC-RB are comparable. The ISAC-DB rate significantly decreases when the vehicle approaches the BS, as a wider beam must be transmitted to cover the entire vehicle. The beam sweeping scheme, due to its limited angle estimation capability, exhibits a generally lower achievable rate and experiences obvious oscillations when the vehicle approaches closer to the BS due to rapid angle variations.

When the transmit antenna size is 32$\times$32, as shown in Fig. \ref{adaptiveVSothernew}(b), the achievable rate of ISAC-RB also decreases as the vehicle approaches the BS. This is due to the narrowing of the HPBW, which causes the beam to widen even when ISAC-RB only needs to cover one CR and one scatterer. The oscillations in the beam sweeping scheme become more pronounced for the same reason.


Additionally, we compare the EKF+MEE algorithm with the Point-target scheme \cite{10659350}, where a scatterer is randomly selected as the ISAC target. For brevity, the results are summarized as the average achievable rate in Table \ref{achievableRateVS}. It is evident that the ISAC-ABA and ISAC-RB schemes significantly outperform the other schemes; however, the ISAC-RB scheme requires substantial computational resources, as detailed in Section VI-C.
\begin{table*}[t]
\caption{The comparisons of average achievable rate using different beamforming strategies\label{achievableRateVS}}
\centering
\renewcommand{\arraystretch}{1.2}
\begin{tabular}{|c|c|c|c|c|c|}
\hline
Transmit antennas & ISAC-ABA (bps/Hz) & ISAC-RB (bps/Hz) & Point-target (bps/Hz) & ISAC-DB (bps/Hz) & Beam sweeping (bps/Hz)\\
\hline
8$\times$8 & 13.04  & 13.05  & 12.47  & 12.65  & 12.39 \\
\hline
16$\times$16  & 14.98 & 15.02 & 14.01 & 14.02 & 13.41\\
\hline
32$\times$32 & 16.96 & 16.86 & 13.72 & 15.06 & 12.81\\
\hline
\end{tabular}
\end{table*}

\begin{table*}[t]
\caption{The comparisons of computational complexity different beamforming strategies\label{complexityAnalysis}}
\centering
\renewcommand{\arraystretch}{1.4}
\begin{tabular}{|c|c|c|c|c|c|c|}
\hline
 & ISAC-ABA ($\Gamma_\text{r1}$=0.04)  & ISAC-ABA ($\Gamma_\text{r1}$=0.02) & ISAC-RB & Point-target  & ISAC-DB  & Beam sweeping \\
\hline
Analytical & ${\cal O}\left( {N_{{\rm{p_\text{num}}}}^3} \right)$  & ${\cal O}\left( {N_{{\rm{p_\text{num}}}}^3} \right)$  & ${\cal O}\left( {N_{{\rm{p_\text{num}}}}^3} \right)$  & ${\cal O}\left( { N_{{\rm{evo}}}^3} \right)$  & ${\cal O}\left( {N_{{\rm{sca}}}^3} \right)$ & ${\cal O}\left( {{N_{{\rm{numb}}}}} \right)$\\
\hline
Total number of calls  & 30 & 70 & 32000 & 32000 & 32000 & 200\\
\hline
\end{tabular}
\end{table*}

\begin{figure}[h] 
\centerline{\includegraphics[width=0.55\textwidth]{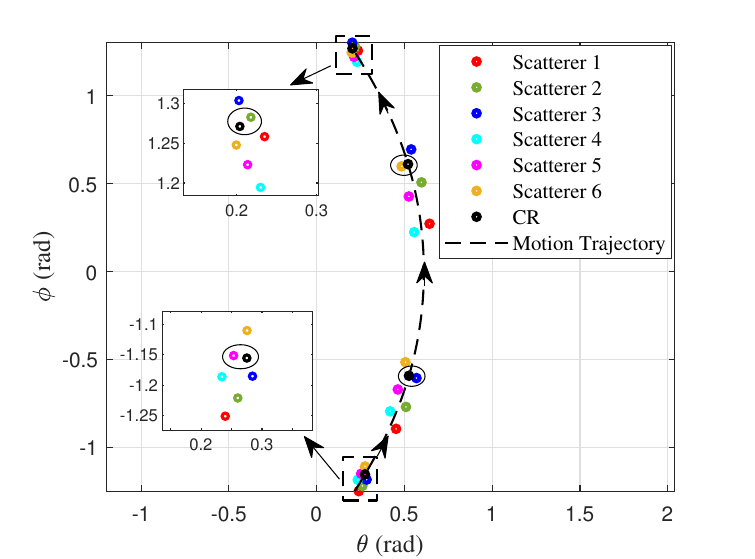}}
\caption{The trajectories of CR and scatterers in the angular domain.}
\label{fig.scaTrajectoryAndEllipseSwitch}
\end{figure}

Furthermore, we simulate the trajectory of the CR and scatterer in the $\theta-\varphi$ domain, as shown in Fig.  \ref{fig.scaTrajectoryAndEllipseSwitch}, where the relative position changes between points is clearly observable.  Note that the relative position here refers to the angular domain rather than the Cartesian coordinate domain. At different times, the scatterer closest to the CR changes, and the BS dynamically switches scatterers to ensure the narrowest beam.


\subsection{Complexity Analysis}
This section provides a complexity analysis of all the schemes listed in Table \ref{achievableRateVS}. For the ISAC-ABA scheme, the MEE algorithm involves three SVD decompositions ${\cal O}\left( {\left( {{N_{{\rm{scacr}}}} + 1} \right)N_{{\rm{svd}}}^3} \right)$, one convex hull calculation ${\cal O}\left( {{N_{{\rm{scacr}}}}{N_{{\rm{p_\text{num}}}}}\log \left( {{N_{{\rm{scacr}}}}{N_{{\rm{p_\text{num}}}}}} \right)} \right)$, two MEE computations ${\cal O}\left( {N_{{\rm{cvh}}}^3 + N_{{\rm{p_\text{num}}}}^3} \right)$, two MEC computations ${\cal O}\left( {N_{{\rm{cvh}}}^3 + N_{{\rm{p_\text{num}}}}^3} \right)$, and one mapping ${\cal O}\left( {N_{{\rm{svd}}}^2{N_{{\rm{cvh}}}}} \right)$, where ${N_{{\rm{scacr}}}} = 2$ denotes the number of input scatterers or CR, ${N_{{\rm{p_\text{num}}}}} = 200$ \footnote{The value of $N_{{\rm{p_\text{num}}}}$ is chosen to ensure that the generated discrete points sufficiently represent the ellipse, thereby guaranteeing the correct operation of the algorithm. In practice, the number of discrete points may be reduced to lower complexity. However, an excessively small number of points may lead to errors in the MEE computation. The same rationale applies to ${N_{\mathrm{cvh}}}$.} represents the number of discrete ellipse points converted from the ellipse parameter, ${N_{{\rm{svd}}}} = 2$ refers to the dimensions of the eigenvalue decomposition matrix, and ${N_{{\rm{cvh}}}}=100$ indicates the number of convex set points. In addition, the time complexity of EKF is ${\cal O}\left( {N_{{\rm{mea}}}^2 + {N_{{\rm{mea}}}}N_{{\rm{evo}}}^2 + N_{{\rm{evo}}}^3} \right)$, where ${N_{{\rm{mea}}}} = 4$ and ${N_{{\rm{evo}}}} = 4$ denote the dimensions of the measurement and the prediction, respectively. Considering that ${\cal O}\left( {N_{{\rm{p_\text{num}}}}^3} \right)$ dominates the overall computational complexity analyzed above, the time complexity of each invocation of the MEE algorithm and the EKF can be considered as ${\cal O}\left( {N_{{\rm{p_\text{num}}}}^3} \right)$, with an average of five executions per invocation. The time complexity of each invocation of ISAC-RB is identical to that of ISAC-ABA. The Point-target scheme exclusively utilizes the EKF, and its time complexity can be considered as ${\cal O}\left( { N_{{\rm{evo}}}^3} \right)$. The ISAC-DB scheme does not utilize the EKF, and its time complexity is ${\cal O}\left( {{N_{{\rm{sca}}}}\log \left( {{N_{{\rm{sca}}}}} \right) + 2N_{{\rm{sca}}}^3 + N_{{\rm{svd}}}^3 + N_{{\rm{svd}}}^2{N_{{\rm{sca}}}}} \right)$, which can be simplified as ${\cal O}\left( {N_{{\rm{sca}}}^3} \right)$, where $N_{{\rm{sca}}}=6$ denotes the number of scatterers. The time complexity of beam sweeping is solely dependent on the number of beams ${N_{{\rm{numb}}}}$, and is expressed as ${\cal O}\left( {{N_{{\rm{numb}}}}} \right)$. Table \ref{complexityAnalysis} summarizes the time complexity and the number of calls for each scheme. Overall, ISAC-ABA achieves nearly the highest achievable rate, while the number of algorithm calls is significantly lower than that of ISAC-RB.

\section{Conclusion}
This paper presented a novel framework for utilizing communication signals to sense motion parameters in V2X networks. Unlike previous studies that primarily relied on qualitative analyses, we derived the CRB for radar parameter estimation tailored to OFDM waveforms and UPA configurations. We developed two NR-V2X-compatible beamforming schemes to address the challenges posed by vehicles modeled as extended targets. The first scheme established beams through a temporally assisted union of predictive error ellipses, enabling accurate scatterer localization. The second scheme introduced an adaptive narrowest-beam strategy during the beam adjustment phase, which effectively tracked the CR while reducing computational complexity. We also addressed the beam design problem using the MEE algorithm combined with practical antenna control techniques. Numerical results verified the feasibility and effectiveness of the proposed approach, demonstrating significant performance gains in realistic V2X scenarios.

The extended target model considered in this paper does not account for the effects of vehicle contours and MPCs. These aspects will be further investigated in future work. Preliminary results addressing this issue can be found in \cite{10494709,9800964,11169427}.

\bibliographystyle{ieeetr}

\end{document}